# Sub-milliarcsecond Imaging of Quasars and AGN

K. I. Kellermann
National Radio Astronomy Observatory, 520 Edgemont Road, Charlottesville, VA 22903*
Electronic mail: kkellerm@nrao.edu

R. C. Vermeulen
Netherlands Foundation for Research in Astronomy, Dwingeloo, The Netherlands
Electronic mail rcv@nfra.nl

J. A. Zensus
Max Planck Institut für Radioastronomie, Auf dem Hugel 69, 53121 Bonn, Germany
Electronic mail: azensus@mpifr-bonn.mpg.de

and

M. H. Cohen
California Institute of Technology, Pasadena, CA 91125
Electronic mail: mhc@astro.caltech.edu

Abstract

We have used the VLBA at 15 GHz to image the structure of 132 strong compact AGN and quasars with a resolution better than one milliarcsecond and a dynamic range typically exceeding 1000 to 1. These observations were made as part of a program to investigate the sub-parsec structure of quasars and AGN and to study the changes in their structure with time. Many of the sources included in our study, particularly those located south of declination +35 degrees, have not been previously imaged with milliarcsecond resolution. Each of the sources has been observed at multiple epochs. In this paper we show images of each of the 132 sources which we have observed. For each source we present data at the epoch which had the best quality data. In a future paper we will discuss the kinematics derived from the observations at all epochs.

Most of the sources we have observed show the canonical core-jet morphology with structure somewhat characteristic of the jets seen on arcsecond scales with the VLA and Westerbork. The milliarcsecond jets generally appear one-sided, but two-sided structure is often found in lower luminosity radio galaxies and in high luminosity quasars with gigahertz peaked spectra. In many cases there is significant curvature, sometimes up to 90 or more, particularly close to the core. In other cases the jets have a more gradual curvature. In some sources there are multiple bends or twists along the jet, suggestive of a three dimensional curved structure. Many of the jets may be described by a small number of apparently discrete components, but in other cases there appears to be a monotonically decreasing distribution of radio emission  Usually the structure is unresolved along the direction perpendicular to the jet, but a few sources have broad plumes. Much of the visible parsec scale structure in compact radio sources can probably be explained as the projection of a relativistic beamed twisted jet, which appears bright at those positions where it approaches the viewer. In some low luminosity radio galaxies, the structure appears more symmetric at 2 cm than at longer wavelengths. The apparent long wavelength asymmetry in these sources is probably due to absorption by intervening ionized material.

A few sources contain only a single component with any secondary feature at least a thousand times weaker. Peak rest frame brightness temperatures are typically of the order of $10^{11\text{-}12}$ K with no evidence for any excess over the limit of $10^{12}$ K expected from inverse Compton cooling. We find no obvious correlation of radio morphology and the detection of gamma-ray emission by EGRET.

*Subject Headings*: galaxies: active – galaxies: nuclei–galaxies: jets

# 1. INTRODUCTION

Earlier VLBI studies have outlined the milliarcsecond structure of hundreds of compact radio sources with a resolution of about 1.5 milliarcsec. The most extensive observations of this type are from the series of Caltech-Jodrell Bank (CJ) surveys made at 1.6 and 5 GHz using antennas in Europe and the United States and covering more than 300 sources north of declination +35 deg (Pearson & Readhead 1981, 1988, Taylor et al. 1994, Polatidis et al. 1995, Thakkar et al. 1995, Xu et al. 1995, Henstock et al. 1995, Taylor et al. 1996).

The milliarcsecond (parsec-scale) structure of quasars and AGN, at radio wavelengths, often appears one-sided with a linear extended feature, usually described as a "jet," e.g., Zensus 1997. In the standard picture (Blandford & Rees 1974), however, the system is intrinsically symmetric, with twin oppositely-directed jets containing relativistic beams. Differential Doppler boosting makes the approaching side appear much brighter than the receding component, and so the objects appear one-sided. A further effect is Doppler favoritism. Objects with jets oriented close to the line-of-sight have their flux density strongly Doppler-boosted, and so they are preferentially found in flux-limited samples of sources chosen at centimeter wavelengths (e.g., Vermeulen & Cohen 1994). A great majority of the sources observed with milliarcsecond resolution show this one-sided morphology, but some objects are two sided. Two-sided morphology can result when the axis is close to the plane of the sky or when the beam velocity is sub-relativistic so that the Doppler boosting is small. Most of these two sided sources have large scale structure, but there is a class of symmetric sources which do not extend beyond 1 kpc. These are referred to as "compact symmetric objects" (CSOs) (e.g., Wilkinson et al. 1994, Readhead et al. 1996a).

In this paper we report on the first part of a source survey being made with the NRAO VLBA (Napier et al. 1994) at a frequency of 15 GHz (2 cm) with a nominal resolution of 0.5 milliarcsecond (mas), (4 parsecs at moderate redshift), in the east-west direction and between 0.6 and 1.3 mas in declination, and with a sensitivity of 200-300 microjanskys rms noise. This is the shortest wavelength and highest resolution which has been used to image a large sample of radio sources. Typically the dynamic range (defined as peak to rms noise) of our images is better than 1000 to 1, but for the weaker sources, it is limited by receiver noise. Throughout this paper we adopt $H_o$ = 65 km/sec/Mpc and $q_0 = 1/2$.

The observations reported here are part of an extended program to study motions and other time variations in source morphology. We show here the structure of each source in our program from one epoch of observation, selected as the one with the best quality data. Images at this and other epochs are displayed on the world wide web at *http://www.cv.nrao.edu/2cmsurvey* and future observations will be included at this site as they become available.

We have exploited the good 2-dimensional coverage of the VLBA to image sources as far south as -20 degree declination with good image quality, and as far south as -40 degrees for a few strong sources. These are the first observations ever made of the sub-arcsecond structure of many of the sources south of declination +35 deg. At declinations north of +35 degrees, many sources have also been observed as part of the CJ 6 and 18 cm surveys.

Our resolution at 2 cm is improved by about a factor of two to three over the Caltech-Jodrell 6 cm global observations. Our 2 cm VLBA images have a resolution comparable to that which will be obtained from the HALCA space VLBI mission at 6 cm, but the 2 cm VLBA images have considerably better sensitivity and dynamic range. Generally, our observing frequency is above the synchrotron self-absorption turnover frequency; thus our observations penetrate further into the nuclear regions than VLBI observations made at lower frequencies.

Our goals were to classify and study the morphological characteristics of compact radio sources; to examine the relation between radio structure and luminosity, spectra, optical counterparts, and redshift; to search for structures that might indicate gravitational minilensing; to find unresolved or nearly unresolved sources to use as VLBA calibrators; and to examine the spectral distribution on parsec scales. In addition to the CJ 6 and 18 cm observations of sources north of declination +35 degrees, many of these sources have also been observed with lower resolution and sensitivity as part of astrometric surveys made at 3.6 and 13 cm (Fey, Clegg, & Fomalont 1996, Fey & Charlot 1997, Peck & Beaseley 1998). Comparison of these longer wavelength images with our 2 cm images will give information on the spectral distribution across individual sources. Also, as the images at these longer wavelengths are generally more sensitive to the larger lower surface brightness structure, comparison of images over a range of wavelengths allows the structure to be traced over a wider range of sizes than is possible at a single VLBA wavelength.

In this paper, we discuss our observing procedures, then present and discuss the images. These images serve as a finding list for those interested in more extensive observing programs. Further analysis and discussion is being deferred to later papers.

## 2. THE SOURCE SAMPLE

Our sample of sources listed in Table 1 is based on the complete 5 GHz 1 Jansky catalogue of Kühr et al. (1981) as supplemented by Stickel, Meisenheimer, & Kühr (1994). Table 1 is organized as follows:

Col 1. IAU source designation.
Col 2. Alternative source name where appropriate.
Col 3 and 4. Right Ascension and Declination (J2000).
Col 5. The optical counterpart denoted as follows: galaxy (G), quasar (Q), BL Lac Object (BL), or Empty Field (EF).
Col 6. Optical magnitude .
Col 7. Redshift.
Col 8. Flux density at 5 GHz taken mostly from Stickel, Meisenheimer, & Kühr (1994).
Col. 9. Indication of whether the source has been detected by the EGRET instrument on GRO (Mattox et al.1997) or not detected (von Montigny et al. 1995b).

Our goal was to include all known sources in the Stickel, Meisenheimer, & Kühr catalogue with spectral index, $\alpha$, flatter than -0.5 ($S \propto \nu^{\alpha}$) at any frequency above 500 MHz and with a measured or extrapolated flux density at 15 GHz greater than 1.5 Jy for sources north of the equator or greater than 2 Jy for sources between declinations -20 deg and the equator. Scheduling constraints did not allow us to include all sources meeting these criteria, but also allowed some weaker sources to be included in the observing schedule. As it turned out, our measurements show that many of the sources we observed had integrated flux densities considerably lower than expected. This is due to variability as well as to uncertainties in the earlier flux density measurements. We have not included known gravitationally lensed sources, except for 0218+357.

In estimating the 15 GHz flux density, we used data taken from VLA measurements listed in the NRAO VLA Calibration Manual, from the University of Michigan observations made available by Aller & Aller (*http://www.astro.lsa.umich.edu/obs/radiotel/umrao.html*), from published spectra (Kühr et al. 1981; Xu et al 1995; Herbig & Readhead, 1995) and unpublished spectra observed by A. Y. Kovalev (private communication) with the RATAN-600 radio telescope, and from the interpolation and extrapolation of measurements at nearby frequencies when 15 GHz observations were not available. Optical counterparts, magnitudes, and redshifts were compiled from a variety of literature sources aided by the NASA Extragalactic Data Base (NED).

## 3. OBSERVING PROCEDURE

The observations were made during eight observing sessions between 1994 and 1997 as summarized in Table 2. In order to optimize the u,v coverage we observed each source over a wide range of hour angle. We deleted data taken from any antenna when its elevation was below ten degrees or during periods of excessively bad weather when the excess sky temperature exceeded 75 degrees, or was changing very rapidly due to heavy rain or thick cloud cover. In Fig. 1, we show the typical u,v coverage for four sources located over a range of declinations.

We observed with a bandwidth 64 MHz (32 MHz in 1994) using one-bit samples and LHC polarization. Each source was observed for four to five minutes once each hour for a range of eight hours (twelve in 1994). Thus we observed three such groups per day, (two in 1994) for a nominal total of 27 sources per day (18 in 1994) at the low (33 kbits/inch) bit density and 30 sources per day at the high (56 kbits/inch) bit density. The rms noise in each image was typically about 400 µJy for the high declination sources observed in 1994 and 200-300 µJy for the other observations, about that expected from the nominal system temperature of 60-100K and antenna gain of about 0.1 K/Jy.

Table 2 shows the log of our observations with the epoch given in column 1, the bit rate in column 2, the linear bit density recorded on tape in kbits/inch in column 3, the number of IF channels in column 4, the approximate total observing time per source in column 5, and the number of sources observed in column 6.

The VLBA uses an FX correlator which gave us between 8 and 256 frequency channels in each of 4-8 IF bands. The differential delays across each IF band and between bands were calibrated by observations of a strong source during each observing session. We then made a conventional global fringe fit to the data, before determining the amplitude and phase of the visibilities. The VLBA is quite stable with time: fringe rates and delays were normally constant to within a few millihertz and a few nanoseconds, respectively. Data from the Hancock and St. Croix sites typically showed larger phase variations due to the significantly greater water vapor column above these antennas.

Initial editing, amplitude calibration using the nominal gain curves determined for each antenna, and fringe fitting were done using the standard NRAO AIPS package. After averaging over frequency, further editing for deviant data points was carried out using the Caltech DIFMAP package, and an automated DIFMAP procedure was used for imaging (Shepherd, Pearson, & Taylor 1994a, 1994b)

Starting with visibility phases self-calibrated with a point source model, the DIFMAP procedure alternates between searches for new CLEAN image components and self calibration. All CLEANing takes place on the residual image, formed from the difference between the observed visibilities and the model as established up to that point. For each CLEAN cycle we made 100 iterations with a loop gain of 0.03. Before each cycle, one new CLEAN window is automatically placed around the highest peak in the current residual image, as long as this peak exceeds 7 to 9 times the current rms image noise. The specific threshold factor was determined experimentally on a few test sources prior to batch processing each observing run, and varied with the data quality. The shape of the rectangular windows is chosen to be an optimal match to the beam shape, and to have an area of 10 to 20 independent beam areas as determined from a few test sources imaged prior to batch processing.

The initial cycles are done using uniform weighting of the data over 2 x 2 u,v grid points. When convergence is reached (no more new peaks meriting a new window and no more flux density to be CLEANed within the windows above 3 times the rms noise), the weighting changes to natural, and the threshold for placing new windows is lowered to 5 to 7 times the residual image rms noise. When this, in turn, converges, self calibration of amplitudes as well as phases is enabled, first by allowing one gain factor per antenna for the entire observation, later by solving for amplitude gains in 30-minute intervals, and finally by having separate solutions for each integration, but only when the peak flux density exceeds the experimentally determined threshold of 0.5 Jy.

With this automated procedure we nearly reached the theoretical noise floor for most sources. For about 25 percent of the sources which had unusually complex structure, were sufficiently strong, or otherwise had inadequate u,v coverage, the above procedure did not result in noise limited images. For these sources we made further iterations of the CLEAN self-cal loop using manually determined parameters.

The resulting images have a nominal resolution of about 0.5 mas along the minor axis, and from 0.6 to 1 mas along the major axis, except for sources near the equator where the major axis is as large as 1.3 mas. Most images are 25.6 mas on a side. In addition to the full resolution images, we examined an image for each source convolved to give a circular beam with resolution of 1.5 milliarcsec. We inspected each of these lower resolution images for structure over an area of 51.2 mas on a side. For sources in which there was evidence for additional emission from other observations, or in which such emission was suggested to be present by our inability to fit the data, we searched a larger area.

### 4. SOURCE STRUCTURE

In Table 3 we summarize the images and give a guide to the contour maps shown in Fig. 2 (a-k). Table 3 is organized as follows:

Col 1.        IAU source designation.
Col 2.        Total flux density in Janskys at 15 GHz measured from our images.
Col 3.        Monochromatic luminosity at 15 GHz in the rest frame of the source calculated assuming a spectral index of zero and isotropic radiation. As we believe most of these sources are relativistically boosted, the true luminosity is probably smaller by a factor of 10 to 100.
Col 4.        Epoch of the VLBA image shown in Fig. 2.
Col 5.        Lowest contour level of the image Fig. 2 which is three times the rms noise in the image.

Col 6 - 8.   Major axis, minor axis, and position angle of the CLEAN beam in mas.
Col 9.   Peak flux density in Janskys per beam measured from the VLBA image.
Col 10.   Peak brightness temperature in the source rest frame calculated from the following expression. $T_b = 7.6 \times 10^{10} S_{Jy} (1+z)/(\theta_{maj}\theta_{min})$ where Sp is the peak flux density in Janskys per beam and $\theta_{maj}$ and $\theta_{min}$ are the semi major and semi minor axis of the beam respectively. In the few cases where no redshift is available, we set z=0.
Col. 11.   Classification of the structure as follows: C (compact), SS (Single-Sided), DS (Double-Sided), or Irr (Irregular) as described below.

In Figs. 2a through 2k, we show a contour map for each source. Figure 3 (Plate 1) shows false color images for six sources indicating the range of structures that are observed.

Pearson and Readhead (1988) have classified the structure of the sources in their pioneering 5 GHz VLBI survey into ten categories and subcategories, most of which exhibited pronounced asymmetry. Later, Wilkinson et al. (1994) called attention to the class of Compact Symmetric Objects (CSOs). More recently, Wilkinson (1995) has described all sources as either core-jets with a wide range of core/jet strengths, jet lengths, and jet bends; or CSOs. Since many of the so-called parsec scale jets do not meet the 4 to 1 length to thickness ratio specified by Bridle and Perley (1984) to define a jet, and so as not to prejudice the interpretation of the underlying physics, we chose to classify each source as either having a compact (C), e.g., barely resolved appearance; a single sided (SS) or double sided (DS) appearance depending on whether the most compact component is located on one side or is in the middle; or an irregular (Irr) 2-dimensional structure.

We note that many of the sources reported here appear to have smooth secondary features or otherwise have complex structures, and we have chosen not to try to describe our images in terms of individual components which may not have any physical reality. Comments on individual sources are given in the next section. We have not made an exhaustive search of the literature to document previous observations of these sources. Rather we comment on peculiarities of the radio structure, observations at other wavelengths, and, where appropriate, other recent published VLBI observations.

Sources which contain only a single barely resolved component are noted as C or "compact." None of these sources are completely unresolved, but they will, nevertheless, be useful as calibrators for future VLBA observations at 2 cm or at nearby wavelengths. In some of these sources, examination of the fringe visibility plots show clear evidence of structure which is not apparent in the images restored with the nominal beamwidth. In the notes to individual sources, where appropriate, we give the structure of these barely resolved sources based on model fitting.

The measured peak brightness temperatures for the most compact sources, as derived from fitting Gaussian components to the u,v data, are in the range of $10^{11-12}$ K. However, only a few sources have an observed brightness temperature which approach the limit of $10^{12}$ K normally expected from a stationary source cooled by Inverse Compton radiation. This is perhaps surprising as it is widely assumed that the most compact feature is the base of a relativistic jet where it turns optically thin, (Blandford & Königl 1979) and that Doppler boosting might be expected to increase the observed brightness temperature beyond the inverse Compton limit. We note also that the nominal brightness temperature limit of $10^{12}$ K is for an idealized homogeneous source with simple geometry (Kellermann & Pauliny-Toth 1969), and that higher brightness temperatures might be observed from more complex geometries.

Using an antenna in space with baselines up to 2.15 Earth diameters at 2.3 GHz, Linfield et al. (1989) reported measured brightness temperatures up to 4 x $10^{12}$ K in apparent excess of the inverse Compton limit. Although it is difficult to measure brightness temperatures above $10^{12}$ K from Earth sized baselines, self-calibration of our ten element array gives a robust calibration that is difficult to obtain from a single distant element, such as from an orbiting antenna. Our sensitivity and calibration are sufficiently good that on our longest baselines near 400 million wavelengths we can determine the fringe visibility to a few percent; and so we can, with confidence, say that a source is resolved if the observed fringe visibility is less than 0.9. This corresponds to a diameter of 0.1 mas or, for a 1 Jy source (with z=0), a brightness temperature of 8 x $10^{11}$ K. We cannot exclude the possibility that some fraction of the observed flux is in an unresolved component embedded in the larger structure, but there is little evidence from examination of our visibility plots, for any unresolved structure which might reflect relativistically boosted jets with apparent brightness temperatures in excess of $10^{12}$ K.

In nearly all of the well resolved sources, there is a single component with dimensions not exceeding that of the synthesized beam and generally located near one of the extremities. These asymmetric structures, which we classify as single sided (SS), are characteristic of the familiar core-jet structure which has been used to describe other milliarcsecond VLBI observations. It is believed that this is the result of differential Doppler boosting of a intrinsic twin jet structure.

In many of the sources with jet-like structure, the jet bends through angles up to 90 degrees or even more. In some cases, the curvature is gradual, but in other sources the jet shows remarkably sharp bends or twists, especially close to the core. Generally, the jet appears unresolved in the direction transverse to its extent, but a few sources have broad plumes. In other sources, the jet is seen to wiggle through several turns of a few tens of degrees in alternate directions. While in some sources the jet appears truly continuous, in others the jet appears to contain multiple more-or-less discrete components. It is not clear if this represents a real distinction, or rather a limited dynamic range of the image.

A few sources show more symmetric double structure or contain three or more components, usually with the central one being the most compact. We refer to these sources as double sided (DS). The nature of the symmetric DS sources is unclear, and it is also not clear if they form a single category. Possible interpretations include a) gravitational minilensing from black holes of the order of $10^6$ $M_\odot$ (e.g., Lacey & Ostriker 1985); b) confined classical doubles (O'Dea, Baum, & Stanghellini 1991); c) young doubles (Fanti et al. 1995, Readhead et al. 1996b); d) reborn doubles (Baum et al. 1990); or e) twin jet sources in the plane of the sky or which move with sub-relativistic velocity. The double sided appearance may also be due to a single jet which, in projection, appears to wrap around both sides of the nucleus. Due to relativistic boosting the jet appears bright first on one side as it approaches the viewer, fades as it bends across the nucleus, and then brightens again as it curves around and approaches the viewer again from the other side.

Both NGC 1052 (0238-084) (Vermeulen et al., in preparation) and Cygnus A (1957+405) show more symmetric structure at 15 GHz than seen at longer wavelengths (cf., Krichbaum et al. 1996, Sorathia et al. 1996). We note that similar phenomena have been reported in other low luminosity radio galaxies such as NGC 1275 (Walker et al. 1998), NGC 4258 (Herrnstein et al. 1997), Hydra A (3C 218) (Taylor 1996), NGC 4261 (3C 270) (Jones & Wehrle 1997), NGC 5128 (Jones et al. 1996), PKS 1413+135 (Perlman et al. 1996), and Mrk 231 (Ulvestad, Wrobel, & Carilli 1998). These high frequency symmetric sources are all identified with galaxies rather than quasars and all have radio luminosities less than $10^{26}$ Watts/Hz. The asymmetry observed on parsec scales at the longer wavelengths in these sources has been interpreted in terms of free-free absorption from a surrounding torus (e.g., Levinson, Laor, & Vermeulen 1995). The surrounding material, in at least some of these sources, is known from HI absorption, and $H_2O$ and OH maser emission, to also have an atomic and molecular component (cf., Conway & Blanco 1995 and references above).

The other sources in our sample which we classify as double sided are 0026+346, 0048-097, 0710+439, 1323+321, 1404+286, and 2021+317. All of these sources, except 0048-097, are included in the (O'Dea, Baum, & Stanghellini 1991) list of Gigahertz Peaked Spectrum (GPS) sources which show a sharp low frequency spectral cutoff near 1 GHz, are identified with galaxies, and have radio luminosity less than about $10^{26}$ Watts/Hz. The data presented by Kühr et al. (1981) indicate that the BL Lac object, 0048-097, as well, is a GPS source. The source 0710+439, has been described as a CSO by Taylor, Readhead, & Pearson (1996) and the other five also appear from our data to have a milliarcsecond morphology characteristic of CSOs. In all of these symmetric or double sided sources, there is no evidence of the effect of differential Doppler boosting, possibly because the source is oriented close to the plane of the sky, or because the motion is non relativistic. However, it is not clear how these powerful CSOs are related to the low luminosity symmetric radio galaxies discussed above, but they too, seem to have a relatively large ambient gas density, as evidenced by recent detections of HI absorption (Vermeulen et al. in preparation).

Other sources in our sample which are classified as GPS sources by O'Dea, Baum, & Stanghellini (1991) are 0153+744, 0316+161, 0528+134, 0552+398, 0615+820, 0738+313, and 2230+114. In addition 2134+004 is a well known GPS source (Shimmins et al. 1968). Except possibly for the optically faint source 0316+161 which does not have a measured redshift, these GPS sources all have a luminosity at 15 GHz much greater than $10^{26}$ Watts/Hz and are identified with quasars rather than galaxies. Earlier lower resolution observations of many GPS sources have been interpreted in terms of compact double structure (e.g., Phillips & Mutel 1982). However, our observations suggest a more diverse distribution of morphologies, at least for the more luminous GPS sources which all appear asymmetric, except 0316+161 which has an irregular structure.

The GPS sources are characterized by a sharp low frequency cutoff in their radio spectra, which might be due to either synchrotron self absorption, free-free absorption in a foreground screen, or to the effects of a dispersive medium (e.g., Kellermann 1966). Considering the apparent asymmetry seen in our images of the more powerful GPS sources, and the corresponding wide range of surface brightness, it is difficult to understand the observed sharp spectral cutoffs of these GPS sources in terms of simple self absorption. Observations of these sources at even shorter wavelengths will be important to see if they appear more symmetric as would be expected if the low frequency cutoff is due to free-free absorption as is found in some of the sources located in galaxies.

Several sources show a complex two dimensional structure which we classify as irregular (Irr). Some of these Irr sources such as 1155+251 and 1611+343 may contain a twisted jet from which we see the relativistically boosted flux from just those regions where the jet is oriented along the line of sight toward the observer. But, we caution that the apparent irregular structure which we observe may simply be the result of inadequate dynamic range and resolution. In particular, our images of these complex sources may be limited by the absence of short spacings in the VLBA.

# 5. NOTES ON INDIVIDUAL SOURCES

**0003-066**: The component about 5 mas to the south is probably an artifact caused by the proximity to the equator.

**0007+106**: There is a probable jet which connects the very compact and highly variable core to a 1 mJy component about 15 mas to the southwest. There may also be a weak component roughly the same distance to the northeast.

**0016+731**: Model fitting to this compact source indicates a double component structure with a separation of 0.7 mas in p.a. -51 and component flux densities of 0.48 Jy and 0.22 Jy.

**0026+346**: We have classified this GPS source as DS assuming that one of the weaker intermediate components is the core of a CSO, but we cannot exclude the possibility that the core is located at one of the extremities.

**0048-097**: This source appears to contain very compact double sided structure.

**0153+744**: The faint components seen to the south east of the brightest component correspond to peaks in a highly curved jet seen by Hummel et al. (1997) at 3.6 and 6 cm.

**0202+149**: The jet shows a nearly right angle bend.

**0218+357**: This is by far the closest spacing lensed image with a component separation of 335 mas (Patnaik, Porcas, & Browne 1995). We do not give a classification for the structure of this source in Table 3 as the appearance is distorted by the lensing.

**0234+285**: There is a very diffuse secondary component.

**0238-084**: This bright elliptical galaxy, NGC 1052 has two jets each containing multiple components moving away from the AGN in opposite directions. The central core is not visible at 2 cm (Vermeulen et al. 1998). Diamond et al. (1997) report two $H_2O$ maser complexes located in the inner part of the western jet.

**0316+162**: CTA 21 is a prototype GPS source (Kellermann et al 1962) and has a complex structure.

**0336-019**: The faint structure to the north may be artifacts.

**0415+379**: While 3C 111 appears clearly one sided at 2 cm, higher resolution 7 mm observations indicate that the apparent double structure seen at the west end of the 2 cm jet breaks up into a more complex multi component symmetric configuration with no obvious component which can be identified as the core (Alef et al. 1998). Our multi-epoch 2 cm observations suggest motion toward the east.

**0430+052**: 3C 120 is a well known superluminal source with a long thin jet with small oscillations. Observations at 18 cm trace the jet out to 200 mas (Benson et al. 1988).

**0458-020**: The jet appears to have a sharp bend within one milliarcsecond of the core. Structure to the south is probably not real.

**0528+134**: This is one of the brightest gamma-ray sources detected by EGRET (Maddox et al. 1997) which has been extensively studied at a number of wavelengths (e.g., Krichbaum et al. 1995).

**0552+398**: Model fitting of this strong compact source indicates a single-sided structure with 3.9 Jy in a small core and 1.1 Jy in a jet component.

**0605-085**: Structure to the west of the brightest component (the core?) could be part of an initial tight loop of a strongly bent jet which continues to the east.

**0615+820**: This source does not have a simple linear morphology.

**0642+449**: The component seen at 6 cm about 3 mas to the east by Xu et al. (1995) is only marginally detected in our 2 cm observations.

**0710+439**: This is a well known CSO (Taylor, Readhead, & Pearson 1996).

**0716+714**: This intraday variable has been observed to show intensity changes up to 10 percent within a day (Wagner et al. 1996).

**0727-115**: The jet appears to have a 90 degree or larger bend.

**0735+178**: The jet has multiple sharp curves.

**0738+313**: The jet has a well defined sharp bend with a prominent knot located at the bend.

**0742+103**: This GPS QSO is unusually faint at optical and IR wavelengths and is among the class of optically quiet quasars (Stickel et al. 1996). Longer wavelength images (Fey & Charlot 1997) show that the structure seen to the north east in our 2 cm image is the beginning of a highly bent jet.

**0814+425**: The jet which emerges to the east bends sharply to the south. The weak feature seen about 6 mas to the south east may be part of the structure reported by Aaron (1996) at 18 cm.

**0859-140**: A long thin curved jet emerges from the core.

**0923+392**: The observed structure at 2 cm appears deceptively simple compared with shorter wavelength observations which show that the western component contains a self absorbed core (Alberdi et al. 1997). Long term monitoring of this source shows that there are both stationary and moving components (Alberdi et al. 1993).

**0945+408**: There is a very diffuse detached secondary component. The inner bright feature has double structure which becomes more noticeable in our observations made the following year.

**1101+384**: Mrk 421 is the closest know BL Lac. It is one of the brightest gamma-ray sources in the sky which has also had a strong TeV flare in May 1994 (Macomb 1995) and again in May 1996 (Zweerink et al. 1997).

**1127-145**: There is a prominent knot at the position of a sharp bend.

**1155+251**: This is an unusually complex source. The jet direction is unclear and there is some indication that the northeastern feature splits into two narrow jets.

**1219+285**: A long thin jet terminates in a prominent extended component.

**1226+023**: The well known jet in 3C 273 continues out to a much further distance (Davis, Unwin, & Muxlow 1991) than shown in our image which is sensitive only to the higher surface brightness structure.

**1228+126**: M87 contains a well known radio jet (e.g., Biretta & Junor 1995). Our observations show that the inner part is well collimated with sharp edges and has an apparent kink about 45 mas away from the core.

**1253-055**: The two bright components in 3C 279 are separating along a direction which differs from the direction defined by the lower surface brightness features. See Carrara et al. (1993) for more details.

**1323+321**: There is no obvious core in this source which is suggestive of a CSO morphology.

**1328+254**: 3C 287 is largely resolved by our observations which show only a weak compact feature. We do not attempt to classify this structure.

**1328+307**: The bright structure seen in our image of 3C 286 is the end of a low surface brightness feature which extends nearly 100 mas to the southwest (Cotton et al. 1997). Cotton et al. suggest that 3C 286 is a normal core-jet source in which the relativistic core (base of the jet) is aimed away from us so we don't see it., but the jet curves into our line of sight. Thus we see a "naked jet."

**1404+286:** From global observations at 6 cm, OQ 208 has been described by Stanghellini et al. (1997) and others as a compact double, characteristic of GHz peaked spectrum sources. Our higher resolution observations at 2 cm, however, show a weak core located between two larger more complex components, suggesting that this source is a CSO. Comparison with the Stanghellini 6 cm image suggests that the central component has an inverted spectrum.

**1413+135**: Multi wavelength observations by Perlman et al. (1996) show complex two sided structure in this source. Our high resolution 2 cm observations, which are insensitive to the low surface brightness steep spectrum northeastern jet, shows the southwestern jet in more detail than seen at the longer wavelengths. 1413+135 is an unusually red quasar which is optically quiet (Rieke, Lebovsky, & Kinman 1979; Stickel et al. 1996).

**1532+016**: The complex structure nearly due south of the core is probably real and suggestive of a twisted jet.

**1611+343**: The complex structure probably reflects a highly twisted jet which, in projection, loops back across itself.

**1641+399**: About three mas from the core, the 3C 345 jet has a sharp transitions to a diffuse jet which is considerably more prominent at longer wavelengths (Lobanov & Zensus 1998).

**1642+690**: The structure and evolution of this source has been described by Venturi et al. (1997) from observations at 3.6 and 6 cm made with lower resolution.

**1652+398**: Mrk 501 has a diffuse twisting jet which has been studied in more detail by Conway & Wrobel (1995). A very high energy (E > 300 Gev) $\gamma$-ray flare was observed in April 1997 (Catanese et al 1997).

**1741-038**: The weak components may be artifacts.

**1845+797**: An unpublished 6 cm image obtained in 1996 by one of us (RCV) shows that the northwestern emission has brightened considerably with respect to the 1982 6 cm image shown by Pearson & Readhead (1988). It is likely that the other, southeastern side, which is the most compact and has the flattest spectrum, is the location of the core.

**1921-293**: This is one of the strongest sources in the sky at millimeter wavelengths and has an unusually flat spectrum between 1 mm and 1 meter. At shorter wavelengths, the spectrum steepens and continues with a spectral index, $\alpha \sim -1$ out to at least ultraviolet wavelengths (Shen et al.1997). The secondary feature is unusually diffuse. VLBA observations made at 7 mm in 1994 and in 1996 (Shen, Moran, & Kellermann, in preparation) show that the jet curves sharply and is elongated along position angle near -23 degrees out to a distance of about 1 mas (4 parsecs) where there is an extension in the 7 mm image toward the diffuse jet which we see at 2 cm located about 6 mas away. Observations at 6 cm made in 1996 are consistent with the 7 mm and 2 cm observations, but 6 cm observations made in 1992 by Shen et al. (1997b) show a secondary feature which lies to the northwest rather than the northeast of the core. Curiously, there is no reported gamma-ray emission from this, one of the brightest known blazars (Mattox et al. 1997).

**1957+405**: The double lobe radio galaxy, Cygnus A, is one of the most luminous radio galaxies known. At 2 cm, the nucleus contains less than one percent of the total luminosity. The two-sided structure seen in our VLBA image is shown more clearly in the images of Sorathia et al. (1996) and Krichbaum et al. (1998) at 0.7, 1.3, and 6 cm.

**2021+614**: This GPS source has been suggested as a likely CSO ( Henstock & Taylor, private communication).

**2131-021**: This object contains a closely spaced double.

**2134+004**: We do not see any evidence for the widely discrepant structures previously reported for this source (Pauliny-Toth et al. 1987, 1989). Much of the reported variations may have been the result of inadequate sampling of the u,v plane and due to the proximity of this source to the equator.

**2145+067**: Model fitting indicates a double source with a component separation of 0.5 mas in p.a. -61 degrees and flux densities of 3.9 Jy and 2.5 Jy.

**2200+420**: The structure of the relatively diffuse jet in BL Lac has been studied in detail by Denn & Mutel et al. (1998).

**2230+114**: CTA 102 is a classical GPS source (Kellermann et al. 1962) containing a jet with multiple twists.

**2251+158**: The jet in 3C 454.3 has pronounced curvature which is also seen in 6 cm observations made between 1981 and 1991 (Pauliny-Toth 1987, 1998).

## 7. GAMMA RAY EMISSION

Observations with the EGRET telescope aboard the Compton Gamma Ray Observatory have shown that strong gamma-ray sources are mostly identified with compact flat spectrum radio sources, particularly those associated with "blazars" and superluminal sources (e.g., Thompson et al. 1993, von Montigny et al. 1995a, Mattox et al. 1997). This is, perhaps, not surprising, as the gamma-ray emission, like the parsec scale radio emission, is thought to originate in a highly relativistic jet (e.g., Dondi & Ghisellini 1995). However, it is not clear why some compact superluminal radio sources are gamma-ray sources and others are not; or more generally, how gamma-ray compact radio sources differ from the non gamma-ray compact radio sources. Curiously, gamma-ray emission has not been detected from some of the most prominent superluminal blazars such as 3C 120 and 3C 345 or from the bright millimeter blazar 1921-293. While the detailed mechanism for the production of gamma-radiation is unclear, it probably takes place close to the central engine (e.g., von Montigny et al. 1995a) or at the base of the relativistic jet (Maddox 1997). Our radio observations probe close to the central engine, but farther out than the likely source of gamma-ray emission. Still we might expect to find differences in the sub-milliarcsecond radio morphology between gamma-ray and non gamma-ray sources, especially if like the radio emission, the gamma-ray emission is beamed and the observed flux density is orientation dependent.

While there may be intrinsic differences among quasars and AGN in their gamma-ray to radio luminosity ratio, the high apparent gamma-ray luminosity combined with the observed rapid variability suggests that the gamma ray emission, like the radio emission, must be beamed. Consideration of relativistic beaming models suggests that there might be a correlation between observed gamma-ray luminosity and the radio morphology, both of which reflect the source orientation. In particular, if the gamma ray emission is more highly beamed than the radio, gamma ray sources would be, on the average, viewed more closely aligned to the line of sight, and this would be reflected in the observed sub-milliarcsecond radio morphology.

Gamma-ray emission has been observed above 100 Mev energy with EGRET for 26 of the sources listed in Table 1 (Mattox et al. 1997). Twenty three other sources in our sample were observed with EGRET, but had no detectable gamma-ray emission (von Montigny et al. 1995b). We find no obvious difference in the radio morphology between these two groups, contrary to what might be expected if the gamma-ray emission is beamed in a more narrow cone than the radio emission, or if there is otherwise a close coupling between the radio and gamma-ray emission.

The absence of any apparent correlation between observed gamma-ray emission and compact structure is perhaps not surprising. Although gamma-ray sources are often identified with bright compact radio sources, the range of gamma-ray fluxes seen by EGRET is only about a factor of 100 and most sources lie within a factor of 10 of the minimum detectable flux. Nearly all EGRET sources are detected only at the time of a strong flare (Mattox et al. 1997). Therefore, it is not unlikely that the other sources in our sample are also gamma-ray emitters, but which are somewhat below the detection level of EGRET or which flared at a time when it was not observed by EGRET. While it will be of interest to obtain complete high resolution radio images for the remaining sources detected by EGRET and to compare their dynamics with those of the non-detections, we do not expect that concentration on the few sources that happened to be above the EGRET detection limit at the time of observation will be of any special interest.

## ACKNOWLEDGMENTS

These observations have depended on the support of many individuals at the NRAO and at Caltech. Our automated DIFMAP procedure was based on scripts written by Martin Shepherd and Greg Taylor. Craig Walker provided up-to-date gain curves for each VLBA antenna as well as the scheduling program. Jon Romney, Peggy Perley and others arranged for the correlation of the data and established the validity of the correlator output. Bill Cotton, Phil Diamond, Eric Greison, Athol Kemball, Leonid Kogan, Pat Murphy and others gave generously of their time to help us through a multitude of problems with implementing new features of the AIPS software package. John Armstrong, an REU student, participated in some of the data reduction in the summer of 1996. We are grateful to all of them for their conscientious support. We also thank the referee, Tim Pearson, for his careful reading of the manuscript and his many constructive suggestions which have helped to improve the paper. RCV was supported in part by NSF grants AST-9117100 and AST-9420018 and MHC by NSF grant AST-9121889. This research has made use of the NASA/IPAC Extragalactic Database (NED), which is operated by the Jet Propulsion Laboratory, Caltech, under contract with the National Aeronautics and Space Administration.

Figure Captions

Figure 1. Plots showing typical (u,v) coverage for survey sources over a range of declinations. a) 1127-145; b) 2134+004; c) 2251+158; d) 1641+399.

Figure 2. a) Contour maps of 12 sources. The lowest contour level is at 3 times the rms noise and is given in column 5 of Table 3. Other contours are shown at increasing powers of two. The peak flux density in each image is given in column 9 of Table 3, and the major axis, minor axis, and position angle of the resorting beam are given in columns 6-8. Most images are centered on the brightest component, but for a few of the larger asymmetric sources, we have shifted the center to fit the image in the 25.6 mas box. Note that while most of the displayed boxes are 25.6 mas on a side, a few of the larger sources are shown in a box twice this size. Each panel also shows a bar representing a linear scale of 10 parsecs for sources with $z > 0.1$ and 1 parsec for sources with $z < 0.1$. b) as in a. The two components of the lensed source 0218+357 are shown in the same panel with the 335 mas separation largely removed; c) as in a; d) as in a; e) as in a; f) as in a; g) as in a; h) as in a; i) as in a; j) as in a); k) as in a.

Figure 3. (Plate 1) Color images of six sources illustrating the range of structures observed. Axes show the offset from the center of the image in milliarcseconds.

TABLE 1. Source List

| Source | Name | R. A. (J2000) | Declination (J2000) | ID | Mag | z | $S_6$ (Jy) | Egret |
|---|---|---|---|---|---|---|---|---|
| (1) | (2) | (3) | (4) | (5) | (6) | (7) | (8) | (9) |
| 0003-066 |         | 0  6 13.893 | -6 23 35.33 | G  | 19.5 | 0.35  | 1.48 |     |
| 0007+106 | IIIZw2  | 0 10 31.007 | 10 58 29.51 | G  | 15.4 | 0.09  | 0.43 |     |
| 0016+731 |         | 0 19 45.787 | 73 27 30.02 | Q  | 18.0 | 1.78  | 1.65 | no  |
| 0026+346 |         | 0 29 14.244 | 34 56 32.26 | G  | 20.7 | 0.52? | 1.27 |     |
| 0035+413 |         | 0 38 24.845 | 41 37  6.00 | Q  | 19.1 | 1.35  | 1.11 |     |
| 0048-097 |         | 0 50 41.317 | -9 29  5.21 | BL | 17.4 | -     | 1.98 |     |
| 0055+300 | NGC315  | 0 57 48.887 | 30 21  8.84 | G  | 12.2 | 0.02  | 1.18 |     |
| 0106+013 |         | 1  8 38.771 |  1 35  0.32 | Q  | 18.3 | 2.10  | 2.28 |     |
| 0112-017 |         | 1 15 17.100 | -1 27  4.58 | Q  | 18.0 | 1.36  | 1.20 |     |
| 0119+041 |         | 1 21 56.862 |  4 22 24.73 | Q  | 19.5 | 0.64  | 1.67 |     |
| 0133+476 | DA55    | 1 36 58.595 | 47 51 29.10 | Q  | 18.0 | 0.86  | 3.22 |     |
| 0149+218 |         | 1 52 18.060 | 22  7  7.70 | Q  | 20.8 | 1.32  | 1.03 |     |
| 0153+744 |         | 1 57 34.967 | 74 42 43.25 | Q  | 16.0 | 2.34  | 1.52 | no  |
| 0202+149 |         | 2  4 50.414 | 15 14 11.04 | G? | 22.1 | -     | 2.47 |     |
| 0202+319 |         | 2  5  4.926 | 32 12 30.10 | Q  | 18.2 | 1.47  | 1.03 |     |
| 0212+735 |         | 2 17 30.817 | 73 49 32.62 | BL | 19.0 | 2.37  | 2.20 | no  |
| 0218+357 |         | 2 21  5.470 | 35 56 13.72 | BL | 20.0 | 1.00  | 1.16 |     |
| 0234+285 | CTD20   | 2 37 52.406 | 28 48  8.99 | Q  | 18.9 | 1.21  | 1.44 | yes |
| 0235+164 |         | 2 38 38.931 | 16 36 59.28 | BL | 19.0 | 0.94  | 2.85 | yes |
| 0238-084 | NGC1052 | 2 41  4.799 | -8 15 20.75 | G  | 11.4 | 0.00  | 1.44 |     |
| 0316+162 | CTA21   | 3 18 57.760 | 16 28 32.34 | Q  | 23.0 | -     | 2.81 |     |
| 0333+321 | NRAO140 | 3 36 30.108 | 32 18 29.34 | Q  | 17.5 | 1.26  | 1.95 | no  |
| 0336-019 | CTA26   | 3 39 30.938 | -1 46 35.80 | Q  | 18.4 | 0.85  | 2.86 | yes |
| 0355+508 | NRAO150 | 3 59 29.748 | 50 57 50.16 | EF | ?    | -     | 3.77 | no  |
| 0415+379 | 3C111   | 4 18 21.277 | 38  1 35.51 | G  | 18.0 | 0.05  | 1.37 | no  |
| 0420-014 |         | 4 23 15.801 | -1 20 33.06 | Q  | 17.8 | 0.92  | 1.46 | yes |
| 0430+052 | 3C120   | 4 33 11.096 |  5 21 15.63 | G  | 14.2 | 0.03  | 8.44 | no  |
| 0440-003 | NRAO190 | 4 42 38.660 |  0 17 43.47 | Q  | 19.2 | 0.84  | 2.61 | yes |
| 0454+844 |         | 5  8 42.363 | 84 32  4.54 | BL | 16.5 | 0.11  | 1.39 | no  |
| 0458-020 |         | 5  1 12.810 | -1 59 14.26 | Q  | 18.4 | 2.29  | 1.74 | yes |
| 0521-365 |         | 5 22 58.012 | -36 27 31.90 | G  | 14.5 | 0.06  | 9.29 | yes |
| 0528+134 |         | 5 30 56.418 | 13 31 55.18 | Q  | 20.0 | 2.06  | 3.97 | yes |
| 0552+398 | DA193   | 5 55 30.806 | 39 48 49.16 | Q  | 18.0 | 2.37  | 5.52 |     |
| 0602+673 |         | 6  7 52.672 | 67 20 55.42 | Q  | 20.6 | 1.97  | 1.06 |     |
| 0605-085 |         | 6  7 59.699 | -8 34 49.98 | Q  | 18.5 | 0.87  | 3.49 | no  |
| 0607-157 |         | 6  9 40.949 | -15 42 40.67 | Q  | 17.0 | 0.32  | 1.82 |     |
| 0615+820 |         | 6 26  3.006 | 82  2 25.57 | Q  | 17.5 | 0.71  | 1.00 | no  |
| 0642+449 | OH471   | 6 46 32.026 | 44 51 16.59 | Q  | 18.5 | 3.41  | 1.22 |     |
| 0707+476 |         | 7 10 46.105 | 47 32 11.14 | Q  | 18.2 | 1.29  | 1.01 |     |
| 0710+439 |         | 7 13 38.177 | 43 49 17.00 | G  | 19.7 | 0.52  | 1.67 |     |
| 0716+714 |         | 7 21 53.449 | 71 20 36.36 | BL | 15.5 | -     | 1.12 | yes |
| 0727-115 |         | 7 30 19.112 | -11 41 12.60 | EF | ?    | -     | 2.2  |     |
| 0735+178 |         | 7 38  7.394 | 17 42 19.00 | BL | 14.8 | 0.42  | 1.99 | yes |
| 0736+017 |         | 7 39 18.034 |  1 37  4.62 | Q  | 16.5 | 0.19  | 1.99 |     |
| 0738+313 |         | 7 41 10.704 | 31 12  0.22 | Q  | 16.1 | 0.63  | 2.49 |     |
| 0742+103 |         | 7 45 33.060 | 10 11 12.69 | EF | ?    | -     | 3.68 |     |

| Source | Name | R. A. (J2000) | Declination (J2000) | ID | Mag | z | $S_6$ (Jy) | Egret |
|---|---|---|---|---|---|---|---|---|
| (1) | (2) | (3) | (4) | (5) | (6) | (7) | (8) | (9) |
| 0745+241 |  | 7 48 36.110 | 24 0 24.15 | Q | 18.5 | 0.41 | 1.01 |  |
| 0748+126 |  | 7 50 52.047 | 12 31 4.83 | Q | 17.8 | 0.89 | 2.28 |  |
| 0754+100 |  | 7 57 6.643 | 9 56 34.85 | BL | 14.5 | 0.66 | 0.90 |  |
| 0804+499 |  | 8 8 39.667 | 49 50 36.53 | Q | 17.5 | 1.43 | 2.05 |  |
| 0808+019 |  | 8 11 26.707 | 1 46 52.22 | BL | 17.5 | - | 1.40 |  |
| 0814+425 |  | 8 18 16.000 | 42 22 45.41 | BL | 18.5 | 0.26 | 1.69 |  |
| 0823+033 |  | 8 25 50.338 | 3 9 24.52 | BL | 18.5 | 0.51 | 1.32 |  |
| 0829+046 |  | 8 31 48.878 | 4 29 39.08 | BL | 16.5 | 0.18 | 1.91 | yes |
| 0850+581 |  | 8 54 41.991 | 57 57 29.95 | Q | 18.0 | 1.32 | 1.39 |  |
| 0851+202 | OJ287 | 8 54 48.875 | 20 6 30.64 | BL | 14.0 | 0.31 | 2.62 |  |
| 0859-140 |  | 9 2 16.831 | -14 15 30.87 | Q | 16.6 | 1.33 | 2.30 |  |
| 0917+449 |  | 9 20 58.459 | 44 41 53.98 | Q | 19.0 | 2.18 | 1.09 |  |
| 0923+392 | 4C39.25 | 9 27 3.014 | 39 2 20.85 | Q | 17.9 | 0.70 | 8.73 | no |
| 0945+408 |  | 9 48 55.339 | 40 39 44.58 | Q | 17.5 | 1.25 | 1.39 |  |
| 0953+254 |  | 9 56 49.876 | 25 15 16.05 | Q | 17.5 | 0.71 | 1.82 |  |
| 1012+232 |  | 10 14 47.067 | 23 1 16.57 | Q | 17.5 | 0.57 | 1.09 |  |
| 1015+359 |  | 10 18 10.988 | 35 42 39.44 | Q | 19.0 | 1.23 | 0.71 |  |
| 1049+215 |  | 10 51 48.790 | 21 19 52.35 | Q | 18.5 | 1.30 | 1.26 |  |
| 1055+018 |  | 10 58 29.605 | 1 33 58.82 | BL | 18.3 | 0.89 | 3.47 |  |
| 1055+201 |  | 10 58 17.902 | 19 51 50.90 | Q | 17.1 | 1.11 | 1.10 |  |
| 1101+384 | Mrk 421 | 11 4 27.315 | 38 12 31.79 | G | 13.3 | 0.03 | 0.72 | yes |
| 1127-145 |  | 11 30 7.053 | -14 49 27.39 | Q | 16.9 | 1.19 | 6.57 | yes |
| 1128+385 |  | 11 30 53.282 | 38 15 18.55 | Q | 18.0 | 1.73 | 0.77 |  |
| 1155+251 |  | 11 58 25.789 | 24 50 18.00 | G | 17.5 | - | 1.16 |  |
| 1156+295 | 4C29.45 | 11 59 31.834 | 29 14 43.82 | Q | 14.4 | 0.73 | 1.46 | yes |
| 1219+285 |  | 12 21 31.691 | 28 13 58.50 | BL | 16.5 | 0.10 | 1.09 | yes |
| 1226+023 | 3C273 | 12 29 6.700 | 2 3 8.60 | Q | 12.9 | 0.16 | 42.80 | yes |
| 1228+126 | M87 | 12 30 49.423 | 12 23 28.04 | G | 9.6 | 0.00 | 74.90 |  |
| 1253-055 | 3C279 | 12 56 11.167 | -5 47 21.52 | Q | 17.8 | 0.54 | 14.90 | yes |
| 1302-102 |  | 13 5 33.015 | -10 33 19.43 | Q | 14.9 | 0.29 | 1.17 |  |
| 1308+326 |  | 13 10 28.664 | 32 20 43.78 | BL | 19.0 | 1.00 | 1.53 |  |
| 1323+321 |  | 13 26 16.514 | 31 54 9.52 | G | 19.0 | 0.37 | 2.28 |  |
| 1328+254 | 3C287 | 13 30 37.690 | 25 9 11.00 | Q | 17.7 | 1.06 | 3.22 |  |
| 1328+307 | 3C286 | 13 31 8.287 | 30 30 32.96 | Q | 17.2 | 0.85 | 7.40 |  |
| 1334-127 |  | 13 37 39.783 | -12 57 24.69 | BL | 17.2 | 0.54 | 2.24 |  |
| 1404+286 | OQ208 | 14 7 0.395 | 28 27 14.69 | G | 16.0 | 0.08 | 2.95 |  |
| 1413+135 |  | 14 15 58.819 | 13 20 23.71 | BL | 20.0 | 0.26 | 0.85 |  |
| 1424+366 |  | 14 26 37.088 | 36 25 9.59 | BS | 18.3 | 1.09 | 0.43 |  |
| 1508-055 |  | 15 10 53.552 | -5 43 6.40 | Q | 17.2 | 1.18 | 2.43 |  |
| 1510-089 |  | 15 12 50.533 | -9 5 59.83 | Q | 16.5 | 0.36 | 3.08 | yes |
| 1532+016 |  | 15 34 52.454 | 1 31 4.21 | Q | 18.7 | 1.44 | 1.14 |  |
| 1546+027 |  | 15 49 29.437 | 2 37 1.16 | Q | 18.0 | 0.41 | 1.45 |  |
| 1548+056 |  | 15 50 35.270 | 5 27 10.47 | Q | 17.7 | 1.42 | 2.24 |  |
| 1606+106 |  | 16 8 46.204 | 10 29 7.78 | Q | 18.5 | 1.23 | 1.49 | yes |

| Source | Name | R. A. (J2000) | Declination (J2000) | ID | Mag | z | $S_6$ (Jy) | Egret |
|---|---|---|---|---|---|---|---|---|
| (1) | (2) | (3) | (4) | (5) | (6) | (7) | (8) | (9) |
| 1611+343 | DA406 | 16 13 41.064 | 34 12 47.91 | Q | 17.5 | 1.40 | 2.67 | yes |
| 1633+382 | | 16 35 15.493 | 38 8 4.50 | Q | 18.0 | 1.81 | 4.02 | yes |
| 1638+398 | NRAO512 | 16 40 29.633 | 39 46 46.03 | Q | 18.5 | 1.66 | 1.15 | |
| 1641+399 | 3C345 | 16 42 58.810 | 39 48 36.99 | Q | 16.0 | 0.59 | 10.80 | no |
| 1642+690 | | 16 42 7.849 | 68 56 39.76 | Q | 19.2 | 0.75 | 1.39 | no |
| 1652+398 | Mrk501 | 16 53 52.217 | 39 45 36.61 | G | 14.2 | 0.03 | 1.42 | |
| 1655+077 | | 16 58 9.011 | 7 41 27.54 | Q | 20.8 | 0.62 | 1.65 | |
| 1656+053 | | 16 58 33.447 | 5 15 16.44 | Q | 16.5 | 0.89 | 2.16 | |
| 1656+477 | | 16 58 2.779 | 47 37 49.24 | Q | 18.0 | 1.62 | 1.24 | |
| 1730-130 | NRAO530 | 17 33 2.706 | -13 4 49.55 | Q | 18.5 | 0.90 | | yes |
| 1741-038 | | 17 43 58.856 | -3 50 4.62 | Q | 18.6 | 1.05 | 3.68 | |
| 1749+096 | | 17 51 32.819 | 9 39 0.73 | BL | 16.8 | 0.32 | 1.87 | |
| 1749+701 | | 17 48 32.840 | 70 5 50.77 | BL | 17.0 | 0.77 | 1.45 | no |
| 1758+388 | | 18 0 24.765 | 38 48 30.70 | Q | 18.0 | 2.09 | 0.74 | |
| 1800+440 | | 18 1 32.315 | 44 4 21.90 | Q | 17.5 | 0.66 | 1.10 | |
| 1803+784 | | 18 0 45.684 | 78 28 4.02 | BL | 17.0 | 0.68 | 2.62 | no |
| 1807+698 | 3C371 | 18 6 50.681 | 69 49 28.11 | BL | 14.4 | 0.05 | 2.26 | no |
| 1823+568 | | 18 24 7.068 | 56 51 1.49 | BL | 18.4 | 0.66 | 1.66 | |
| 1845+797 | 3C390.3 | 18 42 8.990 | 79 46 17.13 | G | 14.4 | 0.06 | 4.38 | no |
| 1901+319 | 3C395 | 19 2 55.935 | 31 59 41.64 | Q | 17.5 | 0.64 | 1.87 | no |
| 1921-293 | OV236 | 19 24 51.056 | -29 14 30.12 | Q | 17.0 | 0.35 | 10.60 | |
| 1928+738 | | 19 27 48.495 | 73 58 1.57 | Q | 16.5 | 0.30 | 3.34 | no |
| 1957+405 | CygA | 19 59 28.348 | 40 44 2.02 | G | ? | 0.06 | 0.75 | |
| 2005+403 | | 20 7 44.945 | 40 29 48.61 | Q | 19.5 | 1.74 | 3.70 | |
| 2007+776 | | 20 5 31.010 | 77 52 43.21 | BL | 16.5 | 0.34 | 1.26 | no |
| 2021+317 | | 20 23 19.018 | 31 53 2.31 | ? | ? | - | 3.05 | |
| 2021+614 | | 20 22 6.682 | 61 36 58.81 | G | 19.5 | 0.23 | 2.31 | |
| 2113+293 | | 21 15 29.414 | 29 33 38.37 | Q | 19.5 | 1.514? | 1.47 | |
| 2131-021 | | 21 34 10.310 | -1 53 17.24 | BL | 18.7 | 0.56 | 2.12 | |
| 2134+004 | | 21 36 38.586 | 0 41 54.20 | Q | 16.8 | 1.93 | 12.30 | no |
| 2136+141 | | 21 39 1.310 | 14 23 35.99 | Q | 18.5 | 2.43 | 1.11 | |
| 2144+092 | | 21 47 10.163 | 9 29 46.68 | Q | 18.9 | 1.11 | 1.01 | |
| 2145+067 | | 21 48 5.459 | 6 57 38.61 | Q | 16.5 | 0.99 | 4.40 | |
| 2200+420 | BLLac | 22 2 43.292 | 42 16 39.98 | BL | 14.5 | 0.07 | 4.77 | yes |
| 2201+315 | | 22 3 14.976 | 31 45 38.27 | Q | 15.5 | 0.30 | 2.32 | |
| 2209+236 | | 22 12 5.968 | 23 55 40.59 | Q | 19.0 | - | 1.12 | yes |
| 2223-052 | 3C446 | 22 25 47.259 | -4 57 1.39 | BL | 17.2 | 1.40 | 4.51 | no |
| 2230+114 | CTA102 | 22 32 36.409 | 11 43 50.89 | Q | 17.3 | 1.04 | 3.61 | yes |
| 2234+282 | | 22 36 22.471 | 28 28 57.42 | Q | 19.0 | 0.80 | 1.06 | |
| 2243-123 | | 22 46 18.232 | -12 6 51.28 | Q | 16.4 | 0.63 | 2.45 | |
| 2251+158 | 3C454.3 | 22 53 57.748 | 16 8 53.56 | Q | 16.1 | 0.86 | 17.40 | yes |
| 2345-167 | | 23 48 2.609 | -16 31 12.02 | Q | 17.5 | 0.58 | 3.36 | |

TABLE 2. Log of the Observations

| Dates | Bit Rate | Linear Bit Density | No. Frequency Channels | Integration Time/Source | No. of Sources |
|---|---|---|---|---|---|
| | Mbit/sec | kbit/in | | min | |
| 1994 Aug. 31 | 64 | 33 | 8 | 60 | 18 |
| 1995 April 4-6 | 128 | 33 | 16 | 30 | 52 |
| 1995 July 28-30 | 128 | 33 | 16 | 30 | 53 |
| 1995 Dec. 15-17 | 128 | 33 | 16 | 30 | 54 |
| 1996 May 18 | 128 | 56 | 16 | 30 | 33 |
| 1996 July 10-11 | 128 | 56 | 16 | 30 | 28 |
| 1996 Oct. 27-29 | 128 | 56 | 16 | 30 | 58 |
| 1997 March 10-14 | 128 | 56 | 256 | 30 | 55 |

TABLE 3. Source Structure

| Source | $S_{total}$ (Jy) | Luminosity (W/Hz) | Epoch | Contour (mJy) | $\theta_{maj}$ (mas) | $\theta_{min}$ (mas) | pa (deg) | $S_{peak}$ (Jy/beam) | $T_b$ (K) | Structure |
|---|---|---|---|---|---|---|---|---|---|---|
| (1) | (2) | (3) | (4) | (5) | (6) | (7) | (8) | (9) | (10) | (11) |
| 0003-066 | 1.92 | $5.1 \times 10^{26}$ | 1996 Oct | 2.1 | 1.1 | 0.5 | -4 | 1.21 | $2.25 \times 10^{10}$ | SS |
| 0007+106 | 0.10 | $1.9 \times 10^{24}$ | 1997 Mar | 0.6 | 1.1 | 0.6 | -1 | 0.10 | $1.21 \times 10^{09}$ | DS |
| 0016+731 | 0.70 | $3.2 \times 10^{27}$ | 1996 Jul | 1.6 | 0.8 | 0.6 | 12 | 0.32 | $1.42 \times 10^{10}$ | C |
| 0026+346 | 0.65 | $3.6 \times 10^{26}$ | 1995 Apr | 0.9 | 0.9 | 0.5 | 12 | 0.10 | $2.61 \times 10^{09}$ | DS |
| 0035+413 | 0.41 | $1.2 \times 10^{27}$ | 1996 Jul | 1.0 | 0.9 | 0.5 | 4 | 0.26 | $1.01 \times 10^{10}$ | SS |
| 0048-097 | 1.34 | - | 1995 Jul | 1.3 | 1.2 | 0.5 | 0 | 1.30 | $1.64 \times 10^{10}$ | C |
| 0055+300 | 0.69 | $4.7 \times 10^{23}$ | 1995 Dec | 0.9 | 0.9 | 0.5 | -6 | 0.35 | $6.00 \times 10^{09}$ | SS |
| 0106+013 | 1.34 | $7.9 \times 10^{27}$ | 1997 Mar | 1.0 | 1.1 | 0.5 | 0 | 1.01 | $4.28 \times 10^{10}$ | SS |
| 0112-017 | 0.82 | $2.4 \times 10^{27}$ | 1995 Jul | 1.4 | 1.2 | 0.5 | 4 | 0.40 | $1.19 \times 10^{10}$ | SS |
| 0119+041 | 1.27 | $1.0 \times 10^{27}$ | 1995 Jul | 1.1 | 1.1 | 0.5 | 0 | 0.98 | $2.20 \times 10^{10}$ | SS |
| 0133+476 | 2.22 | $3.0 \times 10^{27}$ | 1997 Mar | 1.9 | 0.9 | 0.6 | -1 | 1.98 | $5.14 \times 10^{10}$ | SS |
| 0149+218 | 1.42 | $4.0 \times 10^{27}$ | 1995 Dec | 1.2 | 1.1 | 0.5 | 0 | 1.26 | $4.03 \times 10^{10}$ | SS |
| 0153+744 | 0.37 | $2.6 \times 10^{27}$ | 1996 Jul | 1.3 | 0.8 | 0.5 | 29 | 0.19 | $1.22 \times 10^{10}$ | SS |
| 0202+149 | 2.29 | - | 1995 Jul | 2.1 | 1.1 | 0.5 | 1 | 1.80 | $2.47 \times 10^{10}$ | SS |
| 0202+319 | 1.11 | $3.7 \times 10^{27}$ | 1997 Mar | 1.3 | 0.9 | 0.5 | 0 | 1.03 | $4.27 \times 10^{10}$ | SS |
| 0212+735 | 2.69 | $1.9 \times 10^{28}$ | 1994 Aug | 2.3 | 0.7 | 0.6 | -69 | 2.14 | $1.30 \times 10^{11}$ | SS |
| 0218+357 | 1.14 | $2.0 \times 10^{27}$ | 1995 Apr | 1.7 | 0.9 | 0.5 | 1 | 0.44 | $1.47 \times 10^{10}$ | - |
| 0234+285 | 1.56 | $3.8 \times 10^{27}$ | 1995 Dec | 1.6 | 1.0 | 0.5 | 2 | 1.07 | $3.57 \times 10^{10}$ | SS |
| 0235+164 | 0.79 | $1.2 \times 10^{27}$ | 1995 Jul | 1.1 | 1.0 | 0.6 | 1 | 0.67 | $1.64 \times 10^{10}$ | C |
| 0238-084 | 2.08 | $1.3 \times 10^{23}$ | 1997 Mar | 1.0 | 1.2 | 0.5 | 0 | 0.36 | $4.50 \times 10^{09}$ | DS |
| 0316+162 | 0.34 | - | 1997 Mar | 0.8 | 3.0 | 2 | -11 | 0.14 | $1.74 \times 10^{08}$ | Irr |
| 0333+321 | 1.26 | $3.2 \times 10^{27}$ | 1997 Mar | 1.1 | 1.0 | 0.5 | 5 | 0.86 | $2.95 \times 10^{10}$ | SS |
| 0336-019 | 2.23 | $3.0 \times 10^{27}$ | 1997 Mar | 1.0 | 1.2 | 0.5 | 2 | 1.73 | $4.04 \times 10^{10}$ | SS |
| 0355+508 | 3.23 |  | 1995 Apr | 1.2 | 0.8 | 0.5 | 14 | 2.70 | $5.11 \times 10^{10}$ | SS |
| 0415+379 | 5.98 | $3.5 \times 10^{25}$ | 1997 Mar | 4.8 | 0.9 | 0.6 | -3 | 2.28 | $3.35 \times 10^{10}$ | SS |
| 0420-014 | 4.20 | $6.3 \times 10^{27}$ | 1995 Jul | 2.4 | 1.2 | 0.5 | -1 | 3.56 | $8.60 \times 10^{10}$ | SS |
| 0430+052 | 3.01 | $8.2 \times 10^{24}$ | 1997 Mar | 1.9 | 1.1 | 0.5 | 1 | 1.10 | $1.56 \times 10^{10}$ | SS |
| 0440-003 | 1.04 | $1.4 \times 10^{27}$ | 1995 Jul | 1.9 | 1.2 | 0.5 | 2 | 0.61 | $1.41 \times 10^{10}$ | SS |
| 0454+844 | 0.30 | $9.1 \times 10^{24}$ | 1994 Aug | 1.2 | 0.7 | 0.7 | -35 | 0.23 | $3.93 \times 10^{09}$ | SS |
| 0458-020 | 2.33 | $1.6 \times 10^{28}$ | 1995 Jul | 2.5 | 1.2 | 0.6 | 5 | 1.89 | $6.51 \times 10^{10}$ | SS |
| 0521-365 | 2.15 | $1.6 \times 10^{25}$ | 1995 Jul | 1.6 | 3.4 | 0.7 | 19 | 1.52 | $5.10 \times 10^{09}$ | SS |
| 0528+134 | 7.95 | $4.6 \times 10^{28}$ | 1995 Jul | 7.4 | 1.1 | 0.6 | 5 | 6.95 | $2.44 \times 10^{11}$ | SS |
| 0552+398 | 5.02 | $3.6 \times 10^{28}$ | 1997 Mar | 3.1 | 1.0 | 0.5 | 6 | 3.68 | $1.87 \times 10^{11}$ | SS |
| 0602+673 | 0.50 | $2.7 \times 10^{27}$ | 1994 Aug | 1.3 | 0.7 | 0.6 | -45 | 0.42 | $2.25 \times 10^{10}$ | SS |
| 0605-085 | 2.80 | $3.9 \times 10^{27}$ | 1995 Jul | 3.0 | 1.2 | 0.5 | 2 | 2.02 | $4.75 \times 10^{10}$ | SS |
| 0607-157 | 5.60 | $1.3 \times 10^{27}$ | 1995 Jul | 4.9 | 1.3 | 0.5 | 0 | 4.62 | $7.11 \times 10^{10}$ | SS |
| 0615+820 | 0.39 | $3.8 \times 10^{26}$ | 1996 Jul | 1.2 | 0.8 | 0.5 | -25 | 0.17 | $5.43 \times 10^{09}$ | Irr |
| 0642+449 | 2.09 | $2.6 \times 10^{28}$ | 1995 Apr | 2.6 | 0.9 | 0.5 | 5 | 1.73 | $1.28 \times 10^{11}$ | C |
| 0707+476 | 0.41 | $1.1 \times 10^{27}$ | 1996 Jul | 1.2 | 0.8 | 0.5 | -10 | 0.31 | $1.34 \times 10^{10}$ | SS |
| 0710+439 | 0.50 | $2.7 \times 10^{26}$ | 1996 Oct | 0.5 | 0.9 | 0.6 | 22 | 0.11 | $2.32 \times 10^{09}$ | DS |
| 0716+714 | 0.25 | - | 1996 Jul | 0.9 | 0.8 | 0.5 | -16 | 0.23 | $4.29 \times 10^{09}$ | SS |
| 0727-115 | 2.17 | - | 1995 Jul | 2.3 | 1.2 | 0.5 | 1 | 1.62 | $2.05 \times 10^{10}$ | SS |
| 0735+178 | 0.69 | $2.6 \times 10^{26}$ | 1997 Mar | 1.2 | 1.1 | 0.6 | 9 | 0.33 | $5.43 \times 10^{09}$ | SS |
| 0736+017 | 2.58 | $2.2 \times 10^{26}$ | 1995 Jul | 1.4 | 1.2 | 0.5 | 9 | 2.06 | $3.09 \times 10^{10}$ | SS |
| 0738+313 | 1.94 | $1.5 \times 10^{27}$ | 1995 Apr | 1.6 | 1.0 | 0.5 | 8 | 0.80 | $1.97 \times 10^{10}$ | SS |

| Source | $S_{total}$ (Jy) | Luminosity (W/Hz) | Epoch | Contour (mJy) | $\theta_{maj}$ (mas) | $\theta_{min}$ (mas) | pa (deg) | $S_{peak}$ (Jy/beam) | $T_b$ (K) | Structure |
|---|---|---|---|---|---|---|---|---|---|---|
| (1) | (2) | (3) | (4) | (5) | (6) | (7) | (8) | (9) | (10) | (11) |
| 0742+103 | 1.34 | - | 1997 Mar | 1.1 | 1.1 | 0.6 | 10 | 0.51 | $5.82 \times 10^{09}$ | SS |
| 0745+241 | 0.64 | $2.3 \times 10^{26}$ | 1997 Mar | 0.7 | 1.0 | 0.5 | 8 | 0.47 | $9.96 \times 10^{09}$ | SS |
| 0748+126 | 3.25 | $4.6 \times 10^{27}$ | 1997 Mar | 2.1 | 1.1 | 0.6 | 7 | 2.77 | $6.00 \times 10^{10}$ | SS |
| 0754+100 | 1.50 | $1.3 \times 10^{27}$ | 1997 Mar | 1.3 | 1.1 | 0.6 | 1 | 1.12 | $2.12 \times 10^{10}$ | SS |
| 0804+499 | 0.60 | $1.9 \times 10^{27}$ | 1997 Mar | 0.8 | 0.9 | 0.6 | -11 | 0.51 | $1.74 \times 10^{10}$ | SS |
| 0808+019 | 1.34 | - | 1995 Jul | 1.2 | 1.2 | 0.5 | -1 | 1.32 | $1.66 \times 10^{10}$ | C |
| 0814+425 | 0.90 | $1.4 \times 10^{26}$ | 1995 Apr | 0.9 | 0.9 | 0.5 | 12 | 0.54 | $1.15 \times 10^{10}$ | SS |
| 0823+033 | 1.25 | $6.6 \times 10^{26}$ | 1995 Jul | 1.4 | 1.2 | 0.6 | -5 | 0.97 | $1.54 \times 10^{10}$ | SS |
| 0829+046 | 0.97 | $7.4 \times 10^{25}$ | 1996 Oct | 1.0 | 1.0 | 0.4 | -7 | 0.61 | $1.35 \times 10^{10}$ | SS |
| 0850+58 | 0.54 | $1.5 \times 10^{27}$ | 1995 Dec | 1.0 | 0.9 | 0.5 | 17 | 0.30 | $1.19 \times 10^{10}$ | SS |
| 0851+202 | 1.16 | $2.4 \times 10^{26}$ | 1997 Mar | 1.2 | 1.0 | 0.6 | 1 | 0.84 | $1.38 \times 10^{10}$ | SS |
| 0859-140 | 1.58 | $4.5 \times 10^{27}$ | 1995 Jul | 0.8 | 1.2 | 0.5 | 1 | 1.14 | $3.35 \times 10^{10}$ | SS |
| 0917+449 | 1.10 | $6.9 \times 10^{27}$ | 1997 Mar | 0.9 | 0.9 | 0.5 | -1 | 0.70 | $3.75 \times 10^{10}$ | SS |
| 0923+392 | 10.84 | $1.0 \times 10^{28}$ | 1997 Mar | 10.2 | 0.9 | 0.6 | -1 | 7.20 | $1.71 \times 10^{11}$ | SS |
| 0945+408 | 1.37 | $3.5 \times 10^{27}$ | 1995 Apr | 1.4 | 1.1 | 0.5 | -14 | 0.81 | $2.50 \times 10^{10}$ | SS |
| 0953+254 | 1.31 | $1.3 \times 10^{27}$ | 1996 May | 1.7 | 0.9 | 0.5 | -3 | 0.69 | $1.98 \times 10^{10}$ | SS |
| 1012+232 | 1.00 | $6.4 \times 10^{26}$ | 1995 Apr | 1.3 | 1.2 | 0.5 | -12 | 0.76 | $1.49 \times 10^{10}$ | SS |
| 1015+359 | 0.51 | $1.3 \times 10^{27}$ | 1997 Mar | 0.8 | 0.9 | 0.6 | 2 | 0.43 | $1.34 \times 10^{10}$ | SS |
| 1049+215 | 1.04 | $2.8 \times 10^{27}$ | 1997 Mar | 1.0 | 1.0 | 0.5 | 3 | 0.70 | $2.44 \times 10^{10}$ | SS |
| 1055+018 | 2.15 | $3.1 \times 10^{27}$ | 1996 Oct | 3.4 | 1.0 | 0.5 | -3 | 1.57 | $4.49 \times 10^{10}$ | SS |
| 1055+201 | 0.32 | $6.7 \times 10^{26}$ | 1995 Dec | 0.8 | 1.0 | 0.6 | 7 | 0.20 | $5.45 \times 10^{09}$ | SS |
| 1101+384 | 0.52 | $1.2 \times 10^{24}$ | 1997 Mar | 0.7 | 0.9 | 0.5 | 5 | 0.44 | $7.70 \times 10^{09}$ | SS |
| 1127-145 | 2.03 | $4.7 \times 10^{27}$ | 1997 Mar | 1.9 | 1.4 | 0.6 | 7 | 1.00 | $1.98 \times 10^{10}$ | SS |
| 1128+385 | 0.87 | $3.8 \times 10^{27}$ | 1995 Apr | 1.3 | 1.2 | 0.5 | -8 | 0.75 | $2.58 \times 10^{10}$ | SS |
| 1155+251 | 0.24 | - | 1995 Apr | 1.1 | 1.0 | 0.5 | -1 | 0.10 | $1.44 \times 10^{09}$ | Irr |
| 1156+295 | 1.01 | $1.0 \times 10^{27}$ | 1997 Mar | 1.0 | 0.9 | 0.5 | 11 | 0.67 | $1.96 \times 10^{10}$ | SS |
| 1219+285 | 0.43 | $1.1 \times 10^{25}$ | 1995 Apr | 1.2 | 1.0 | 0.5 | -5 | 0.20 | $3.35 \times 10^{09}$ | SS |
| 1226+023 | 25.72 | $1.5 \times 10^{27}$ | 1997 Mar | 8.6 | 1.1 | 0.5 | 0 | 9.18 | $1.46 \times 10^{11}$ | SS |
| 1228+126 | 2.40 | $1.1 \times 10^{23}$ | 1997 Mar | 0.8 | 1.1 | 0.6 | 0 | 1.12 | $1.29 \times 10^{10}$ | SS |
| 1253-055 | 21.56 | $1.3 \times 10^{28}$ | 1997 Mar | 4.8 | 1.2 | 0.5 | 1 | 15.43 | $2.99 \times 10^{11}$ | SS |
| 1302-102 | 0.70 | $1.3 \times 10^{26}$ | 1995 Jul | 1.3 | 1.2 | 0.5 | 0 | 0.54 | $8.72 \times 10^{09}$ | SS |
| 1308+326 | 3.31 | $5.8 \times 10^{27}$ | 1996 May | 4.9 | 0.9 | 0.5 | -8 | 2.09 | $7.02 \times 10^{10}$ | C |
| 1323+321 | 0.65 | $1.9 \times 10^{26}$ | 1995 Apr | 0.8 | 1.1 | 0.4 | 20 | 0.04 | $9.42 \times 10^{08}$ | DS |
| 1328+254 | 0.08 | $1.5 \times 10^{26}$ | 1995 Apr | 2.3 | 0.9 | 0.4 | 2 | 0.09 | $3.76 \times 10^{09}$ | - |
| 1328+307 | 0.78 | $1.0 \times 10^{27}$ | 1997 Mar | 1.4 | 2.6 | 1.7 | 21 | 0.17 | $5.28 \times 10^{08}$ | Irr |
| 1334-127 | 5.10 | $3.0 \times 10^{27}$ | 1995 Jul | 3.7 | 1.3 | 0.5 | 1 | 4.64 | $8.31 \times 10^{10}$ | SS |
| 1404+286 | 0.97 | $1.4 \times 10^{25}$ | 1995 Dec | 1.2 | 1.0 | 0.5 | 1 | 0.61 | $9.95 \times 10^{09}$ | DS |
| 1413+135 | 1.58 | $2.4 \times 10^{26}$ | 1995 Jul | 1.6 | 1.0 | 0.5 | 3 | 1.41 | $2.68 \times 10^{10}$ | SS |
| 1424+366 | 0.43 | $8.7 \times 10^{26}$ | 1997 Mar | 0.7 | 1.0 | 0.6 | 13 | 0.30 | $7.93 \times 10^{09}$ | SS |
| 1508-055 | 0.73 | $1.7 \times 10^{27}$ | 1997 Mar | 0.7 | 1.2 | 0.5 | 7 | 0.54 | $1.49 \times 10^{10}$ | SS |
| 1510-089 | 1.20 | $3.4 \times 10^{26}$ | 1995 Jul | 2.5 | 1.4 | 0.6 | 13 | 0.74 | $9.09 \times 10^{09}$ | SS |
| 1532+016 | 0.76 | $2.4 \times 10^{27}$ | 1997 Mar | 0.8 | 1.1 | 0.5 | 6 | 0.36 | $1.20 \times 10^{10}$ | Irr |
| 1546+027 | 2.83 | $1.0 \times 10^{27}$ | 1996 Oct | 3.2 | 1.0 | 0.5 | -2 | 2.64 | $5.63 \times 10^{10}$ | SS |
| 1548+056 | 2.83 | $8.9 \times 10^{27}$ | 1997 Mar | 0.9 | 1.1 | 0.5 | 6 | 2.10 | $6.98 \times 10^{10}$ | SS |
| 1606+106 | 1.56 | $3.9 \times 10^{27}$ | 1997 Mar | 1.1 | 1.1 | 0.5 | 8 | 1.23 | $3.77 \times 10^{10}$ | SS |

| Source | $S_{total}$ (Jy) | Luminosity (W/Hz) | Epoch | Contour (mJy) | $\theta_{maj}$ (mas) | $\theta_{min}$ (mas) | pa (deg) | $S_{peak}$ (Jy/beam) | $T_b$ (K) | Structure |
|---|---|---|---|---|---|---|---|---|---|---|
| (1) | (2) | (3) | (4) | (5) | (6) | (7) | (8) | (9) | (10) | (11) |
| 1611+343 | 4.05 | $1.2 \times 10^{28}$ | 1997 Mar | 1.9 | 0.9 | 0.5 | 17 | 2.72 | $1.10 \times 10^{11}$ | Irr |
| 1633+382 | 1.40 | $6.5 \times 10^{27}$ | 1997 Mar | 1.5 | 1.0 | 0.6 | -6 | 0.73 | $2.59 \times 10^{10}$ | SS |
| 1638+398 | 0.98 | $4.0 \times 10^{27}$ | 1995 Apr | 1.1 | 0.9 | 0.5 | -6 | 0.78 | $3.49 \times 10^{10}$ | C |
| 1641+399 | 8.48 | $5.9 \times 10^{27}$ | 1995 Apr | 2.3 | 0.9 | 0.5 | -5 | 4.53 | $1.21 \times 10^{11}$ | SS |
| 1642+690 | 0.63 | $6.7 \times 10^{26}$ | 1996 Jul | 1.1 | 0.7 | 0.5 | 3 | 0.36 | $1.38 \times 10^{10}$ | SS |
| 1652+398 | 0.77 | $2.2 \times 10^{24}$ | 1997 Mar | 0.4 | 0.9 | 0.5 | -7 | 0.38 | $6.59 \times 10^{09}$ | SS |
| 1655+077 | 1.73 | $1.3 \times 10^{27}$ | 1997 Mar | 1.6 | 1.2 | 0.5 | -4 | 1.40 | $2.87 \times 10^{10}$ | SS |
| 1656+053 | 0.46 | $6.6 \times 10^{26}$ | 1996 Jul | 1.6 | 1.2 | 0.6 | -5 | 0.26 | $5.25 \times 10^{09}$ | SS |
| 1656+477 | 1.05 | $4.1 \times 10^{27}$ | 1997 Mar | 0.9 | 0.9 | 0.5 | -7 | 0.66 | $2.90 \times 10^{10}$ | SS |
| 1730-130 | 9.88 | $1.4 \times 10^{28}$ | 1995 Jul | 3.4 | 1.4 | 0.6 | 3 | 8.79 | $1.51 \times 10^{11}$ | SS |
| 1741-038 | 4.06 | $7.8 \times 10^{27}$ | 1995 Jul | 2.7 | 1.3 | 0.6 | 6 | 3.77 | $7.52 \times 10^{10}$ | C |
| 1749+096 | 5.58 | $1.3 \times 10^{27}$ | 1995 Jul | 3.5 | 1.2 | 0.6 | 10 | 5.49 | $7.63 \times 10^{10}$ | SS |
| 1749+701 | 0.55 | $6.1 \times 10^{26}$ | 1997 Mar | 0.7 | 0.8 | 0.5 | -5 | 0.40 | $1.35 \times 10^{10}$ | SS |
| 1758+388 | 1.43 | $8.4 \times 10^{27}$ | 1995 Apr | 1.6 | 0.9 | 0.5 | 3 | 1.22 | $6.34 \times 10^{10}$ | SS |
| 1800+440 | 1.02 | $8.7 \times 10^{26}$ | 1997 Mar | 0.9 | 0.9 | 0.5 | -1 | 0.93 | $2.60 \times 10^{10}$ | SS |
| 1803+784 | 2.05 | $1.8 \times 10^{27}$ | 1997 Mar | 1.1 | 0.8 | 0.5 | -5 | 1.46 | $4.63 \times 10^{10}$ | SS |
| 1807+698 | 1.13 | $7.3 \times 10^{24}$ | 1997 Mar | 0.8 | 0.8 | 0.5 | -3 | 0.66 | $1.31 \times 10^{10}$ | SS |
| 1823+568 | 2.31 | $2.0 \times 10^{27}$ | 1995 Dec | 1.4 | 0.9 | 0.5 | 3 | 2.05 | $5.73 \times 10^{10}$ | SS |
| 1845+797 | 0.37 | $2.9 \times 10^{24}$ | 1997 Mar | 0.6 | 0.8 | 0.5 | 4 | 0.18 | $3.57 \times 10^{09}$ | SS |
| 1901+319 | 1.09 | $8.6 \times 10^{26}$ | 1997 Mar | 1.1 | 1.0 | 0.6 | -10 | 0.64 | $1.33 \times 10^{10}$ | SS |
| 1921-293 | 14.39 | $3.9 \times 10^{27}$ | 1995 Jul | 14.4 | 1.8 | 0.6 | 17 | 9.57 | $9.06 \times 10^{10}$ | SS |
| 1928+738 | 3.04 | $6.2 \times 10^{26}$ | 1996 Oct | 3.2 | 0.7 | 0.5 | 16 | 1.15 | $3.24 \times 10^{10}$ | SS |
| 1957+405 | 1.68 | $1.3 \times 10^{25}$ | 1995 Dec | 1.5 | 1.0 | 0.5 | 9 | 0.60 | $9.52 \times 10^{09}$ | DS |
| 2005+403 | 2.51 | $1.1 \times 10^{28}$ | 1997 Mar | 2.8 | 0.9 | 0.5 | -8 | 1.02 | $4.69 \times 10^{10}$ | SS |
| 2007+777 | 1.05 | $2.7 \times 10^{26}$ | 1995 Dec | 1.7 | 0.8 | 0.5 | -14 | 0.61 | $1.55 \times 10^{10}$ | SS |
| 2021+317 | 2.02 | - | 1996 May | 3.3 | 0.9 | 0.5 | -14 | 1.13 | $1.90 \times 10^{10}$ | DS |
| 2021+614 | 2.21 | $2.6 \times 10^{26}$ | 1995 Dec | 2.6 | 0.9 | 0.5 | -8 | 1.19 | $2.45 \times 10^{10}$ | SS |
| 2113+293 | 0.94 | $3.3 \times 10^{27}$ | 1995 Apr | 2.1 | 0.9 | 0.5 | -1 | 0.87 | $3.66 \times 10^{10}$ | SS |
| 2131-021 | 1.22 | $7.6 \times 10^{26}$ | 1995 Jul | 1.3 | 1.5 | 0.7 | 9 | 0.83 | $9.30 \times 10^{09}$ | SS |
| 2134+004 | 5.51 | $2.9 \times 10^{28}$ | 1995 Jul | 4.3 | 1.3 | 0.6 | 9 | 2.21 | $6.30 \times 10^{10}$ | SS |
| 2136+141 | 1.92 | $1.4 \times 10^{28}$ | 1995 Jul | 2.0 | 1.2 | 0.6 | 14 | 1.45 | $5.21 \times 10^{10}$ | SS |
| 2144+092 | 0.56 | $1.2 \times 10^{27}$ | 1995 Jul | 1.4 | 1.2 | 0.6 | 13 | 0.44 | $9.79 \times 10^{09}$ | SS |
| 2145+067 | 6.52 | $1.1 \times 10^{28}$ | 1997 Mar | 7.9 | 1.1 | 0.6 | 3 | 3.70 | $8.45 \times 10^{10}$ | C |
| 2200+420 | 3.23 | $3.7 \times 10^{25}$ | 1997 Mar | 3.2 | 1.0 | 0.6 | 7 | 1.49 | $2.00 \times 10^{10}$ | SS |
| 2201+315 | 3.10 | $6.1 \times 10^{26}$ | 1997 Mar | 2.4 | 1.0 | 0.5 | 5 | 2.28 | $4.48 \times 10^{10}$ | SS |
| 2209+236 | 0.91 | - | 1995 Apr | 2.3 | 0.9 | 0.5 | 6 | 0.80 | $1.35 \times 10^{10}$ | SS |
| 2223-052 | 3.92 | $1.2 \times 10^{28}$ | 1997 Mar | 2.3 | 1.2 | 0.5 | 5 | 3.28 | $9.93 \times 10^{10}$ | SS |
| 2230+114 | 2.33 | $4.3 \times 10^{27}$ | 1995 Jul | 2.5 | 1.3 | 0.7 | 16 | 0.81 | $1.37 \times 10^{10}$ | SS |
| 2234+282 | 1.21 | $1.4 \times 10^{27}$ | 1995 Apr | 2.1 | 0.9 | 0.5 | 2 | 0.74 | $2.25 \times 10^{10}$ | SS |
| 2243-123 | 2.56 | $2.0 \times 10^{27}$ | 1997 Mar | 1.9 | 1.2 | 0.5 | 0 | 1.85 | $3.80 \times 10^{10}$ | SS |
| 2251+158 | 8.86 | $1.2 \times 10^{28}$ | 1997 Mar | 5.5 | 1.1 | 0.6 | 10 | 4.24 | $9.02 \times 10^{10}$ | SS |
| 2345-167 | 1.60 | $1.1 \times 10^{27}$ | 1997 Mar | 1.8 | 1.2 | 0.5 | 4 | 0.75 | $1.50 \times 10^{10}$ | SS |

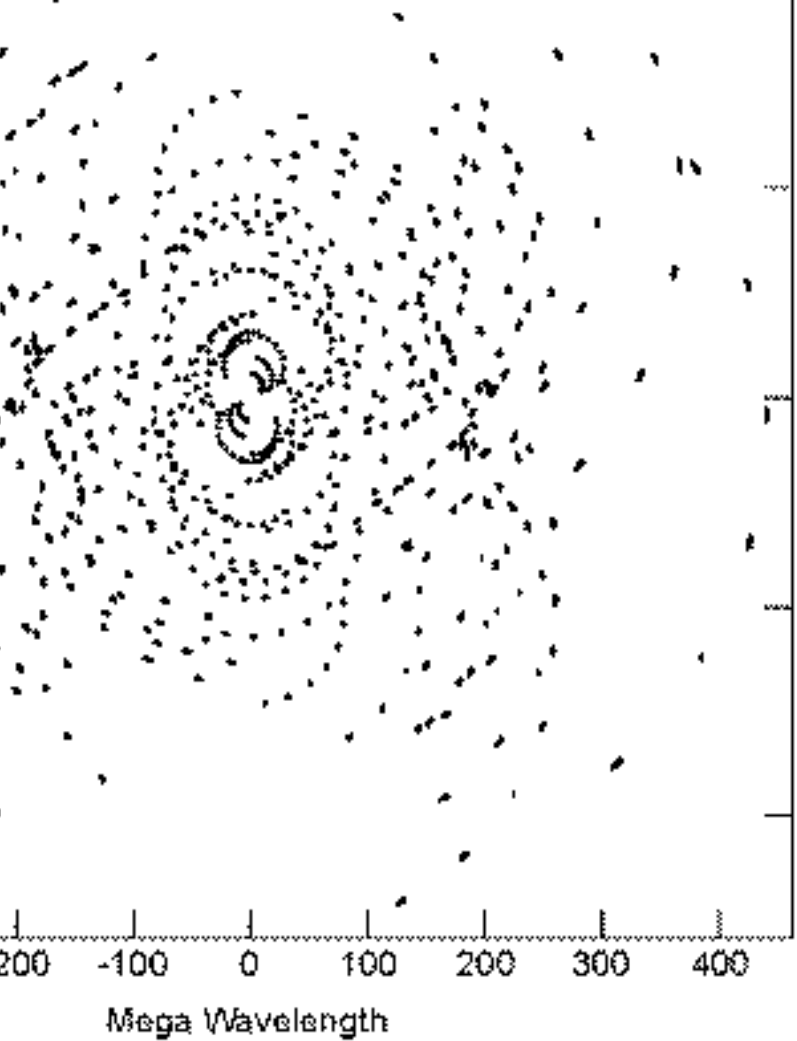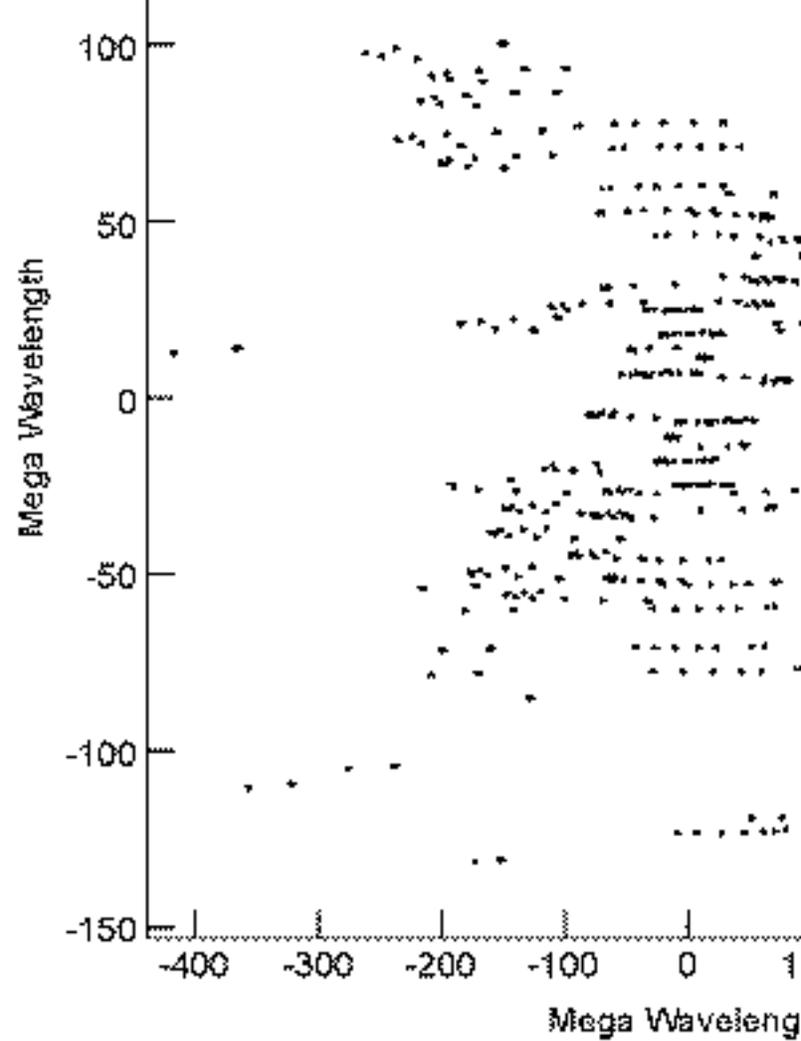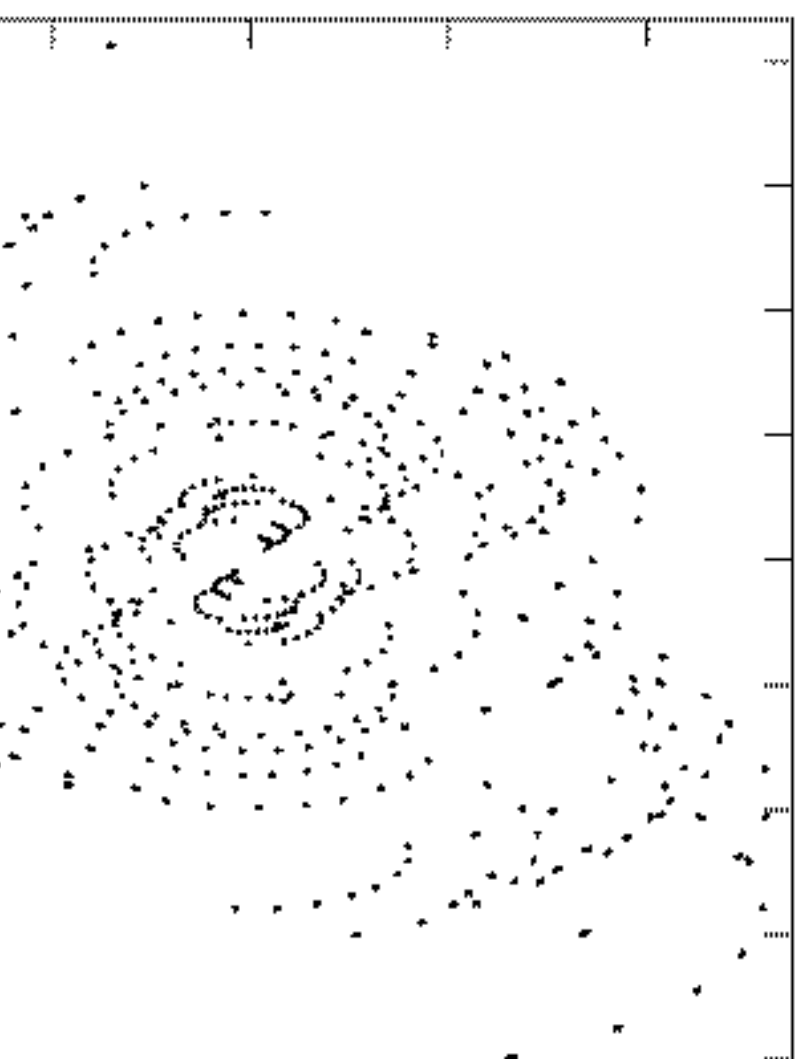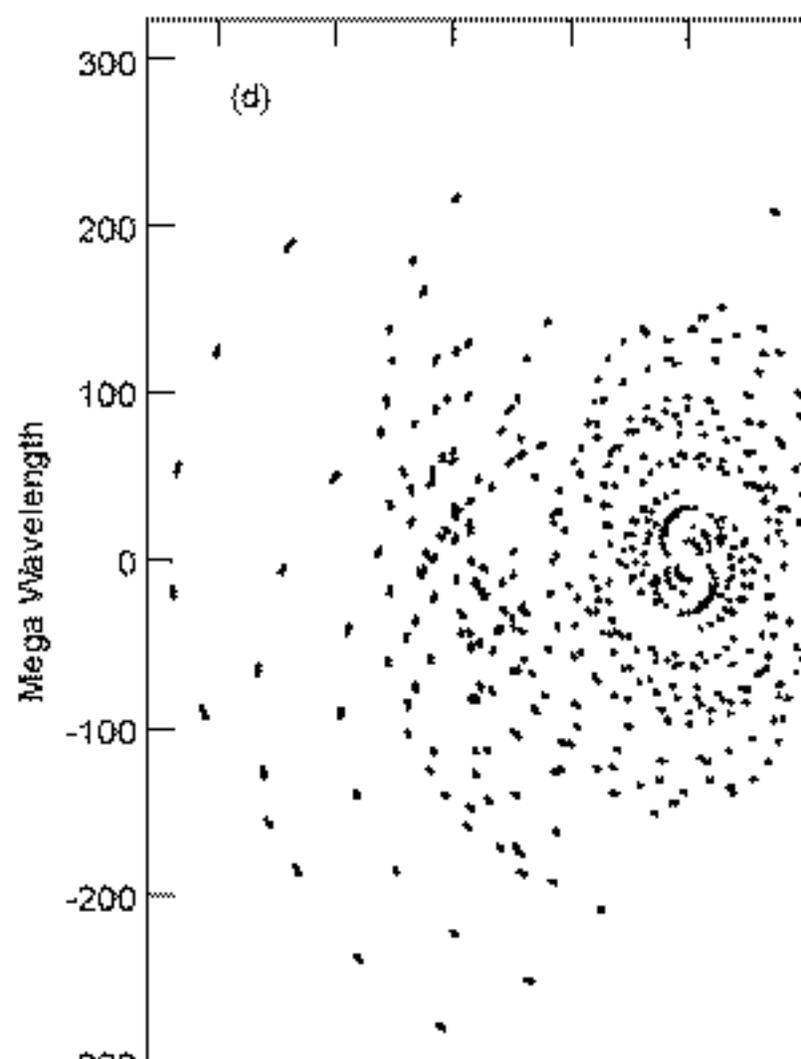

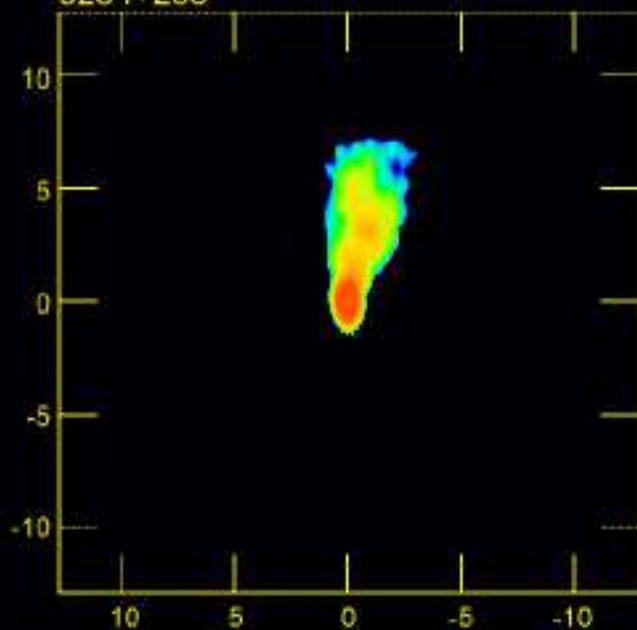
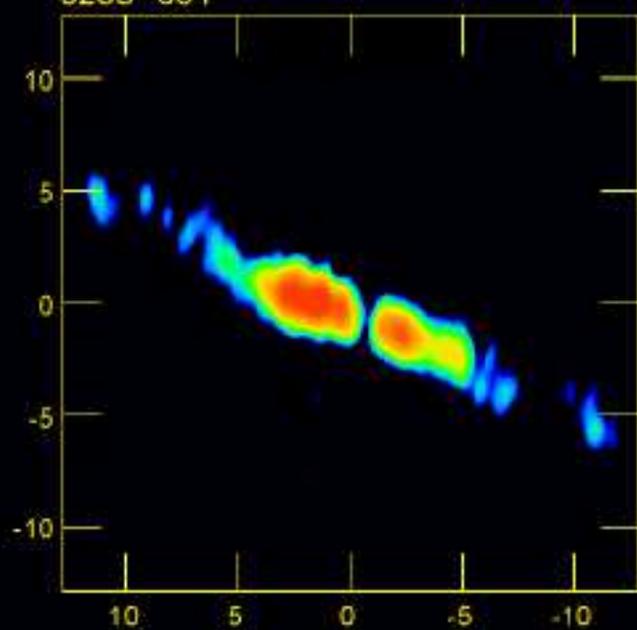
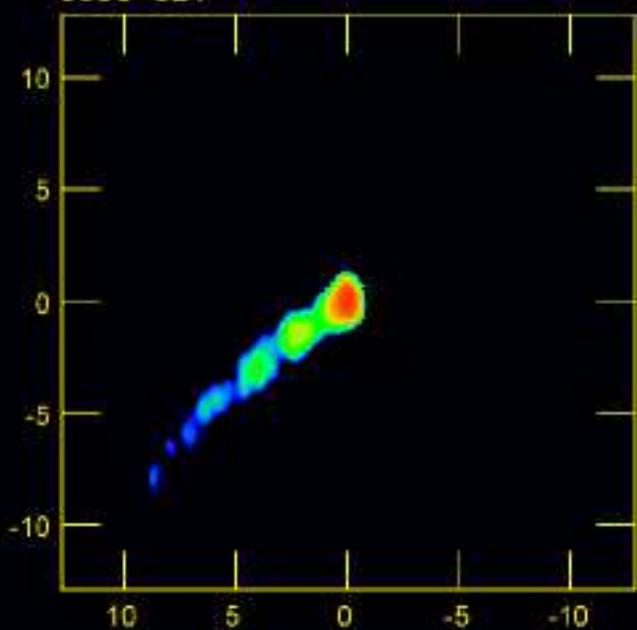
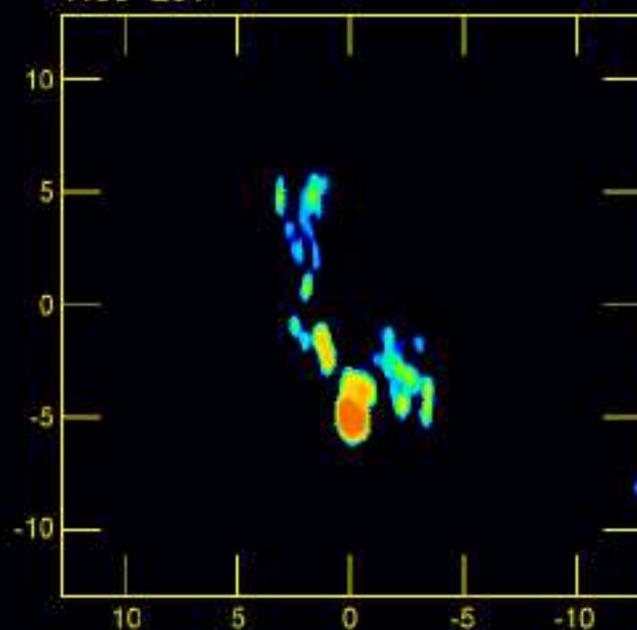
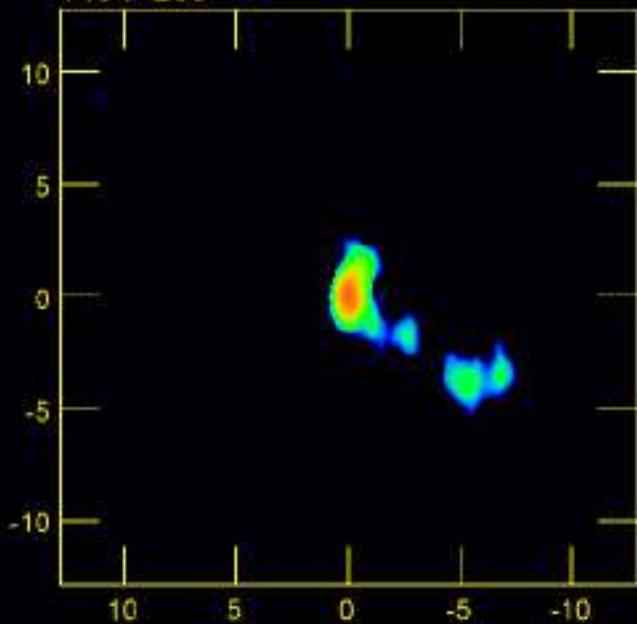
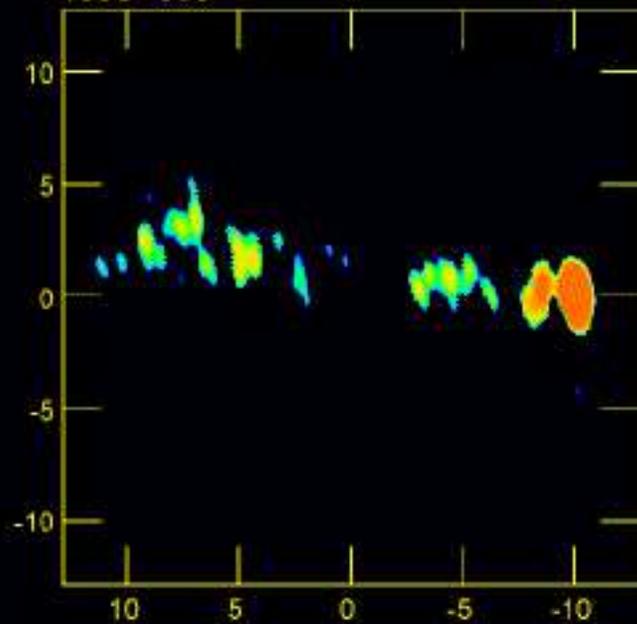

TABLE 3. Source Structure

| Source | $S_{total}$ (Jy) | Luminosity (W/Hz) | Epoch | Contour (mJy) | $\theta_{maj}$ (mas) | $\theta_{min}$ (mas) | pa (deg) | $S_{peak}$ (Jy/beam) | $T_b$ (K) | Structure |
|---|---|---|---|---|---|---|---|---|---|---|
| (1) | (2) | (3) | (4) | (5) | (6) | (7) | (8) | (9) | (10) | (11) |
| 0003-066 | 1.92 | $5.1 \times 10^{26}$ | 1996 Oct | 2.1 | 1.1 | 0.5 | -4 | 1.21 | $2.25 \times 10^{10}$ | SS |
| 0007+106 | 0.10 | $1.9 \times 10^{24}$ | 1997 Mar | 0.6 | 1.1 | 0.6 | -1 | 0.10 | $1.21 \times 10^{09}$ | DS |
| 0016+731 | 0.70 | $3.2 \times 10^{27}$ | 1996 Jul | 1.6 | 0.8 | 0.6 | 12 | 0.32 | $1.42 \times 10^{10}$ | C |
| 0026+346 | 0.65 | $3.6 \times 10^{26}$ | 1995 Apr | 0.9 | 0.9 | 0.5 | 12 | 0.10 | $2.61 \times 10^{09}$ | DS |
| 0035+413 | 0.41 | $1.2 \times 10^{27}$ | 1996 Jul | 1.0 | 0.9 | 0.5 | 4 | 0.26 | $1.01 \times 10^{10}$ | SS |
| 0048-097 | 1.34 | - | 1995 Jul | 1.3 | 1.2 | 0.5 | 0 | 1.30 | $1.64 \times 10^{10}$ | C |
| 0055+300 | 0.69 | $4.7 \times 10^{23}$ | 1995 Dec | 0.9 | 0.9 | 0.5 | -6 | 0.35 | $6.00 \times 10^{09}$ | SS |
| 0106+013 | 1.34 | $7.9 \times 10^{27}$ | 1997 Mar | 1.0 | 1.1 | 0.5 | 0 | 1.01 | $4.28 \times 10^{10}$ | SS |
| 0112-017 | 0.82 | $2.4 \times 10^{27}$ | 1995 Jul | 1.4 | 1.2 | 0.5 | 4 | 0.40 | $1.19 \times 10^{10}$ | SS |
| 0119+041 | 1.27 | $1.0 \times 10^{27}$ | 1995 Jul | 1.1 | 1.1 | 0.5 | 0 | 0.98 | $2.20 \times 10^{10}$ | SS |
| 0133+476 | 2.22 | $3.0 \times 10^{27}$ | 1997 Mar | 1.9 | 0.9 | 0.6 | -1 | 1.98 | $5.14 \times 10^{10}$ | SS |
| 0149+218 | 1.42 | $4.0 \times 10^{27}$ | 1995 Dec | 1.2 | 1.1 | 0.5 | 0 | 1.26 | $4.03 \times 10^{10}$ | SS |
| 0153+744 | 0.37 | $2.6 \times 10^{27}$ | 1996 Jul | 1.3 | 0.8 | 0.5 | 29 | 0.19 | $1.22 \times 10^{10}$ | SS |
| 0202+149 | 2.29 | - | 1995 Jul | 2.1 | 1.1 | 0.5 | 1 | 1.80 | $2.47 \times 10^{10}$ | SS |
| 0202+319 | 1.11 | $3.7 \times 10^{27}$ | 1997 Mar | 1.3 | 0.9 | 0.5 | 0 | 1.03 | $4.27 \times 10^{10}$ | SS |
| 0212+735 | 2.69 | $1.9 \times 10^{28}$ | 1994 Aug | 2.3 | 0.7 | 0.6 | -69 | 2.14 | $1.30 \times 10^{11}$ | SS |
| 0218+357 | 1.14 | $2.0 \times 10^{27}$ | 1995 Apr | 1.7 | 0.9 | 0.5 | 1 | 0.44 | $1.47 \times 10^{10}$ | - |
| 0234+285 | 1.56 | $3.8 \times 10^{27}$ | 1995 Dec | 1.6 | 1.0 | 0.5 | 2 | 1.07 | $3.57 \times 10^{10}$ | SS |
| 0235+164 | 0.79 | $1.2 \times 10^{27}$ | 1995 Jul | 1.1 | 1.0 | 0.6 | 1 | 0.67 | $1.64 \times 10^{10}$ | C |
| 0238-084 | 2.08 | $1.3 \times 10^{23}$ | 1997 Mar | 1.0 | 1.2 | 0.5 | 0 | 0.36 | $4.50 \times 10^{09}$ | DS |
| 0316+162 | 0.34 | - | 1997 Mar | 0.8 | 3.0 | 2 | -11 | 0.14 | $1.74 \times 10^{08}$ | Irr |
| 0333+321 | 1.26 | $3.2 \times 10^{27}$ | 1997 Mar | 1.1 | 1.0 | 0.5 | 5 | 0.86 | $2.95 \times 10^{10}$ | SS |
| 0336-019 | 2.23 | $3.0 \times 10^{27}$ | 1997 Mar | 1.0 | 1.2 | 0.5 | 2 | 1.73 | $4.04 \times 10^{10}$ | SS |
| 0355+508 | 3.23 | | 1995 Apr | 1.2 | 0.8 | 0.5 | 14 | 2.70 | $5.11 \times 10^{10}$ | SS |
| 0415+379 | 5.98 | $3.5 \times 10^{25}$ | 1997 Mar | 4.8 | 0.9 | 0.6 | -3 | 2.28 | $3.35 \times 10^{10}$ | SS |
| 0420-014 | 4.20 | $6.3 \times 10^{27}$ | 1995 Jul | 2.4 | 1.2 | 0.5 | -1 | 3.56 | $8.60 \times 10^{10}$ | SS |
| 0430+052 | 3.01 | $8.2 \times 10^{24}$ | 1997 Mar | 1.9 | 1.1 | 0.5 | 1 | 1.10 | $1.56 \times 10^{10}$ | SS |
| 0440-003 | 1.04 | $1.4 \times 10^{27}$ | 1995 Jul | 1.9 | 1.2 | 0.5 | 2 | 0.61 | $1.41 \times 10^{10}$ | SS |
| 0454+844 | 0.30 | $9.1 \times 10^{24}$ | 1994 Aug | 1.2 | 0.7 | 0.7 | -35 | 0.23 | $3.93 \times 10^{09}$ | SS |
| 0458-020 | 2.33 | $1.6 \times 10^{28}$ | 1995 Jul | 2.5 | 1.2 | 0.6 | 5 | 1.89 | $6.51 \times 10^{10}$ | SS |
| 0521-365 | 2.15 | $1.6 \times 10^{25}$ | 1995 Jul | 1.6 | 3.4 | 0.7 | 19 | 1.52 | $5.10 \times 10^{09}$ | SS |
| 0528+134 | 7.95 | $4.6 \times 10^{28}$ | 1995 Jul | 7.4 | 1.1 | 0.6 | 5 | 6.95 | $2.44 \times 10^{11}$ | SS |
| 0552+398 | 5.02 | $3.6 \times 10^{28}$ | 1997 Mar | 3.1 | 1.0 | 0.5 | 6 | 3.68 | $1.87 \times 10^{11}$ | SS |
| 0602+673 | 0.50 | $2.7 \times 10^{27}$ | 1994 Aug | 1.3 | 0.7 | 0.6 | -45 | 0.42 | $2.25 \times 10^{10}$ | SS |
| 0605-085 | 2.80 | $3.9 \times 10^{27}$ | 1995 Jul | 3.0 | 1.2 | 0.5 | 2 | 2.02 | $4.75 \times 10^{10}$ | SS |
| 0607-157 | 5.60 | $1.3 \times 10^{27}$ | 1995 Jul | 4.9 | 1.3 | 0.5 | 0 | 4.62 | $7.11 \times 10^{10}$ | SS |
| 0615+820 | 0.39 | $3.8 \times 10^{26}$ | 1996 Jul | 1.2 | 0.8 | 0.5 | -25 | 0.17 | $5.43 \times 10^{09}$ | Irr |
| 0642+449 | 2.09 | $2.6 \times 10^{28}$ | 1995 Apr | 2.6 | 0.9 | 0.5 | 5 | 1.73 | $1.28 \times 10^{11}$ | C |
| 0707+476 | 0.41 | $1.1 \times 10^{27}$ | 1996 Jul | 1.2 | 0.8 | 0.5 | -10 | 0.31 | $1.34 \times 10^{10}$ | SS |
| 0710+439 | 0.50 | $2.7 \times 10^{26}$ | 1996 Oct | 0.5 | 0.9 | 0.6 | 22 | 0.11 | $2.32 \times 10^{09}$ | DS |
| 0716+714 | 0.25 | - | 1996 Jul | 0.9 | 0.8 | 0.5 | -16 | 0.23 | $4.29 \times 10^{09}$ | SS |
| 0727-115 | 2.17 | - | 1995 Jul | 2.3 | 1.2 | 0.5 | 1 | 1.62 | $2.05 \times 10^{10}$ | SS |
| 0735+178 | 0.69 | $2.6 \times 10^{26}$ | 1997 Mar | 1.2 | 1.1 | 0.6 | 9 | 0.33 | $5.43 \times 10^{09}$ | SS |
| 0736+017 | 2.58 | $2.2 \times 10^{26}$ | 1995 Jul | 1.4 | 1.2 | 0.5 | 9 | 2.06 | $3.09 \times 10^{10}$ | SS |
| 0738+313 | 1.94 | $1.5 \times 10^{27}$ | 1995 Apr | 1.6 | 1.0 | 0.5 | 8 | 0.80 | $1.97 \times 10^{10}$ | SS |
| 0742+103 | 1.34 | - | 1997 Mar | 1.1 | 1.1 | 0.6 | 10 | 0.51 | $5.82 \times 10^{09}$ | SS |
| 0745+241 | 0.64 | $2.3 \times 10^{26}$ | 1997 Mar | 0.7 | 1.0 | 0.5 | 8 | 0.47 | $9.96 \times 10^{09}$ | SS |

| Source | $S_{total}$ (Jy) | Luminosity (W/Hz) | Epoch | Contour (mJy) | $\theta_{maj}$ (mas) | $\theta_{min}$ (mas) | pa (deg) | $S_{peak}$ (Jy/beam) | $T_b$ (K) | Structure |
|---|---|---|---|---|---|---|---|---|---|---|
| (1) | (2) | (3) | (4) | (5) | (6) | (7) | (8) | (9) | (10) | (11) |
| 0748+126 | 3.25 | $4.6 \times 10^{27}$ | 1997 Mar | 2.1 | 1.1 | 0.6 | 7 | 2.77 | $6.00 \times 10^{10}$ | SS |
| 0754+100 | 1.50 | $1.3 \times 10^{27}$ | 1997 Mar | 1.3 | 1.1 | 0.6 | 1 | 1.12 | $2.12 \times 10^{10}$ | SS |
| 0804+499 | 0.60 | $1.9 \times 10^{27}$ | 1997 Mar | 0.8 | 0.9 | 0.6 | -11 | 0.51 | $1.74 \times 10^{10}$ | SS |
| 0808+019 | 1.34 | - | 1995 Jul | 1.2 | 1.2 | 0.5 | -1 | 1.32 | $1.66 \times 10^{10}$ | C |
| 0814+425 | 0.90 | $1.4 \times 10^{26}$ | 1995 Apr | 0.9 | 0.9 | 0.5 | 12 | 0.54 | $1.15 \times 10^{10}$ | SS |
| 0823+033 | 1.25 | $6.6 \times 10^{26}$ | 1995 Jul | 1.4 | 1.2 | 0.6 | -5 | 0.97 | $1.54 \times 10^{10}$ | SS |
| 0829+046 | 0.97 | $7.4 \times 10^{25}$ | 1996 Oct | 1.0 | 1.0 | 0.4 | -7 | 0.61 | $1.35 \times 10^{10}$ | SS |
| 0850+58 | 0.54 | $1.5 \times 10^{27}$ | 1995 Dec | 1.0 | 0.9 | 0.5 | 17 | 0.30 | $1.19 \times 10^{10}$ | SS |
| 0851+202 | 1.16 | $2.4 \times 10^{26}$ | 1997 Mar | 1.2 | 1.0 | 0.6 | 1 | 0.84 | $1.38 \times 10^{10}$ | SS |
| 0859-140 | 1.58 | $4.5 \times 10^{27}$ | 1995 Jul | 0.8 | 1.2 | 0.5 | 1 | 1.14 | $3.35 \times 10^{10}$ | SS |
| 0917+449 | 1.10 | $6.9 \times 10^{27}$ | 1997 Mar | 0.9 | 0.9 | 0.5 | -1 | 0.70 | $3.75 \times 10^{10}$ | SS |
| 0923+392 | 10.84 | $1.0 \times 10^{28}$ | 1997 Mar | 10.2 | 0.9 | 0.6 | -1 | 7.20 | $1.71 \times 10^{11}$ | SS |
| 0945+408 | 1.37 | $3.5 \times 10^{27}$ | 1995 Apr | 1.4 | 1.1 | 0.5 | -14 | 0.81 | $2.50 \times 10^{10}$ | SS |
| 0953+254 | 1.31 | $1.3 \times 10^{27}$ | 1996 May | 1.7 | 0.9 | 0.5 | -3 | 0.69 | $1.98 \times 10^{10}$ | SS |
| 1012+232 | 1.00 | $6.4 \times 10^{26}$ | 1995 Apr | 1.3 | 1.2 | 0.5 | -12 | 0.76 | $1.49 \times 10^{10}$ | SS |
| 1015+359 | 0.51 | $1.3 \times 10^{27}$ | 1997 Mar | 0.8 | 0.9 | 0.6 | 2 | 0.43 | $1.34 \times 10^{10}$ | SS |
| 1049+215 | 1.04 | $2.8 \times 10^{27}$ | 1997 Mar | 1.0 | 1.0 | 0.5 | 3 | 0.70 | $2.44 \times 10^{10}$ | SS |
| 1055+018 | 2.15 | $3.1 \times 10^{27}$ | 1996 Oct | 3.4 | 1.0 | 0.5 | -3 | 1.57 | $4.49 \times 10^{10}$ | SS |
| 1055+201 | 0.32 | $6.7 \times 10^{26}$ | 1995 Dec | 0.8 | 1.0 | 0.6 | 7 | 0.20 | $5.45 \times 10^{09}$ | SS |
| 1101+384 | 0.52 | $1.2 \times 10^{24}$ | 1997 Mar | 0.7 | 0.9 | 0.5 | 5 | 0.44 | $7.70 \times 10^{09}$ | SS |
| 1127-145 | 2.03 | $4.7 \times 10^{27}$ | 1997 Mar | 1.9 | 1.4 | 0.6 | 7 | 1.00 | $1.98 \times 10^{10}$ | SS |
| 1128+385 | 0.87 | $3.8 \times 10^{27}$ | 1995 Apr | 1.3 | 1.2 | 0.5 | -8 | 0.75 | $2.58 \times 10^{10}$ | SS |
| 1155+251 | 0.24 | - | 1995 Apr | 1.1 | 1.0 | 0.5 | -1 | 0.10 | $1.44 \times 10^{09}$ | Irr |
| 1156+295 | 1.01 | $1.0 \times 10^{27}$ | 1997 Mar | 1.0 | 0.9 | 0.5 | 11 | 0.67 | $1.96 \times 10^{10}$ | SS |
| 1219+285 | 0.43 | $1.1 \times 10^{25}$ | 1995 Apr | 1.2 | 1.0 | 0.5 | -5 | 0.20 | $3.35 \times 10^{09}$ | SS |
| 1226+023 | 25.72 | $1.5 \times 10^{27}$ | 1997 Mar | 8.6 | 1.1 | 0.5 | 0 | 9.18 | $1.46 \times 10^{11}$ | SS |
| 1228+126 | 2.40 | $1.1 \times 10^{23}$ | 1997 Mar | 0.8 | 1.1 | 0.6 | 0 | 1.12 | $1.29 \times 10^{10}$ | SS |
| 1253-055 | 21.56 | $1.3 \times 10^{28}$ | 1997 Mar | 4.8 | 1.2 | 0.5 | 1 | 15.43 | $2.99 \times 10^{11}$ | SS |
| 1302-102 | 0.70 | $1.3 \times 10^{26}$ | 1995 Jul | 1.3 | 1.2 | 0.5 | 0 | 0.54 | $8.72 \times 10^{09}$ | SS |
| 1308+326 | 3.31 | $5.8 \times 10^{27}$ | 1996 May | 4.9 | 0.9 | 0.5 | -8 | 2.09 | $7.02 \times 10^{10}$ | C |
| 1323+321 | 0.65 | $1.9 \times 10^{26}$ | 1995 Apr | 0.8 | 1.1 | 0.4 | 20 | 0.04 | $9.42 \times 10^{08}$ | DS |
| 1328+254 | 0.08 | $1.5 \times 10^{26}$ | 1995 Apr | 2.3 | 0.9 | 0.4 | 2 | 0.09 | $3.76 \times 10^{09}$ | - |
| 1328+307 | 0.78 | $1.0 \times 10^{27}$ | 1997 Mar | 1.4 | 2.6 | 1.7 | 21 | 0.17 | $5.28 \times 10^{08}$ | Irr |
| 1334-127 | 5.10 | $3.0 \times 10^{27}$ | 1995 Jul | 3.7 | 1.3 | 0.5 | 1 | 4.64 | $8.31 \times 10^{10}$ | SS |
| 1404+286 | 0.97 | $1.4 \times 10^{25}$ | 1995 Dec | 1.2 | 1.0 | 0.5 | 1 | 0.61 | $9.95 \times 10^{09}$ | DS |
| 1413+135 | 1.58 | $2.4 \times 10^{26}$ | 1995 Jul | 1.6 | 1.0 | 0.5 | 3 | 1.41 | $2.68 \times 10^{10}$ | SS |
| 1424+366 | 0.43 | $8.7 \times 10^{26}$ | 1997 Mar | 0.7 | 1.0 | 0.6 | 13 | 0.30 | $7.93 \times 10^{09}$ | SS |
| 1508-055 | 0.73 | $1.7 \times 10^{27}$ | 1997 Mar | 0.7 | 1.2 | 0.5 | 7 | 0.54 | $1.49 \times 10^{10}$ | SS |
| 1510-089 | 1.20 | $3.4 \times 10^{26}$ | 1995 Jul | 2.5 | 1.4 | 0.6 | 13 | 0.74 | $9.09 \times 10^{09}$ | SS |
| 1532+016 | 0.76 | $2.4 \times 10^{27}$ | 1997 Mar | 0.8 | 1.1 | 0.5 | 6 | 0.36 | $1.20 \times 10^{10}$ | Irr |
| 1546+027 | 2.83 | $1.0 \times 10^{27}$ | 1996 Oct | 3.2 | 1.0 | 0.5 | -2 | 2.64 | $5.63 \times 10^{10}$ | SS |
| 1548+056 | 2.83 | $8.9 \times 10^{27}$ | 1997 Mar | 0.9 | 1.1 | 0.5 | 6 | 2.10 | $6.98 \times 10^{10}$ | SS |
| 1606+106 | 1.56 | $3.9 \times 10^{27}$ | 1997 Mar | 1.1 | 1.1 | 0.5 | 8 | 1.23 | $3.77 \times 10^{10}$ | SS |

| Source | $S_{total}$ (Jy) | Luminosity (W/Hz) | Epoch | Contour (mJy) | $\theta_{maj}$ (mas) | $\theta_{min}$ (mas) | pa (deg) | $S_{peak}$ (Jy/beam) | $T_b$ (K) | Structure |
|---|---|---|---|---|---|---|---|---|---|---|
| (1) | (2) | (3) | (4) | (5) | (6) | (7) | (8) | (9) | (10) | (11) |
| 1611+343 | 4.05 | $1.2 \times 10^{28}$ | 1997 Mar | 1.9 | 0.9 | 0.5 | 17 | 2.72 | $1.10 \times 10^{11}$ | Irr |
| 1633+382 | 1.40 | $6.5 \times 10^{27}$ | 1997 Mar | 1.5 | 1.0 | 0.6 | -6 | 0.73 | $2.59 \times 10^{10}$ | SS |
| 1638+398 | 0.98 | $4.0 \times 10^{27}$ | 1995 Apr | 1.1 | 0.9 | 0.5 | -6 | 0.78 | $3.49 \times 10^{10}$ | C |
| 1641+399 | 8.48 | $5.9 \times 10^{27}$ | 1995 Apr | 2.3 | 0.9 | 0.5 | -5 | 4.53 | $1.21 \times 10^{11}$ | SS |
| 1642+690 | 0.63 | $6.7 \times 10^{26}$ | 1996 Jul | 1.1 | 0.7 | 0.5 | 3 | 0.36 | $1.38 \times 10^{10}$ | SS |
| 1652+398 | 0.77 | $2.2 \times 10^{24}$ | 1997 Mar | 0.4 | 0.9 | 0.5 | -7 | 0.38 | $6.59 \times 10^{09}$ | SS |
| 1655+077 | 1.73 | $1.3 \times 10^{27}$ | 1997 Mar | 1.6 | 1.2 | 0.5 | -4 | 1.40 | $2.87 \times 10^{10}$ | SS |
| 1656+053 | 0.46 | $6.6 \times 10^{26}$ | 1996 Jul | 1.6 | 1.2 | 0.6 | -5 | 0.26 | $5.25 \times 10^{09}$ | SS |
| 1656+477 | 1.05 | $4.1 \times 10^{27}$ | 1997 Mar | 0.9 | 0.9 | 0.5 | -7 | 0.66 | $2.90 \times 10^{10}$ | SS |
| 1730-130 | 9.88 | $1.4 \times 10^{28}$ | 1995 Jul | 3.4 | 1.4 | 0.6 | 3 | 8.79 | $1.51 \times 10^{11}$ | SS |
| 1741-038 | 4.06 | $7.8 \times 10^{27}$ | 1995 Jul | 2.7 | 1.3 | 0.6 | 6 | 3.77 | $7.52 \times 10^{10}$ | C |
| 1749+096 | 5.58 | $1.3 \times 10^{27}$ | 1995 Jul | 3.5 | 1.2 | 0.6 | 10 | 5.49 | $7.63 \times 10^{10}$ | SS |
| 1749+701 | 0.55 | $6.1 \times 10^{26}$ | 1997 Mar | 0.7 | 0.8 | 0.5 | -5 | 0.40 | $1.35 \times 10^{10}$ | SS |
| 1758+388 | 1.43 | $8.4 \times 10^{27}$ | 1995 Apr | 1.6 | 0.9 | 0.5 | 3 | 1.22 | $6.34 \times 10^{10}$ | SS |
| 1800+440 | 1.02 | $8.7 \times 10^{26}$ | 1997 Mar | 0.9 | 0.9 | 0.5 | -1 | 0.93 | $2.60 \times 10^{10}$ | SS |
| 1803+784 | 2.05 | $1.8 \times 10^{27}$ | 1997 Mar | 1.1 | 0.8 | 0.5 | -5 | 1.46 | $4.63 \times 10^{10}$ | SS |
| 1807+698 | 1.13 | $7.3 \times 10^{24}$ | 1997 Mar | 0.8 | 0.8 | 0.5 | -3 | 0.66 | $1.31 \times 10^{10}$ | SS |
| 1823+568 | 2.31 | $2.0 \times 10^{27}$ | 1995 Dec | 1.4 | 0.9 | 0.5 | 3 | 2.05 | $5.73 \times 10^{10}$ | SS |
| 1845+797 | 0.37 | $2.9 \times 10^{24}$ | 1997 Mar | 0.6 | 0.8 | 0.5 | 4 | 0.18 | $3.57 \times 10^{09}$ | SS |
| 1901+319 | 1.09 | $8.6 \times 10^{26}$ | 1997 Mar | 1.1 | 1.0 | 0.6 | -10 | 0.64 | $1.33 \times 10^{10}$ | SS |
| 1921-293 | 14.39 | $3.9 \times 10^{27}$ | 1995 Jul | 14.4 | 1.8 | 0.6 | 17 | 9.57 | $9.06 \times 10^{10}$ | SS |
| 1928+738 | 3.04 | $6.2 \times 10^{26}$ | 1996 Oct | 3.2 | 0.7 | 0.5 | 16 | 1.15 | $3.24 \times 10^{10}$ | SS |
| 1957+405 | 1.68 | $1.3 \times 10^{25}$ | 1995 Dec | 1.5 | 1.0 | 0.5 | 9 | 0.60 | $9.52 \times 10^{09}$ | DS |
| 2005+403 | 2.51 | $1.1 \times 10^{28}$ | 1997 Mar | 2.8 | 0.9 | 0.5 | -8 | 1.02 | $4.69 \times 10^{10}$ | SS |
| 2007+777 | 1.05 | $2.7 \times 10^{26}$ | 1995 Dec | 1.7 | 0.8 | 0.5 | -14 | 0.61 | $1.55 \times 10^{10}$ | SS |
| 2021+317 | 2.02 | - | 1996 May | 3.3 | 0.9 | 0.5 | -14 | 1.13 | $1.90 \times 10^{10}$ | DS |
| 2021+614 | 2.21 | $2.6 \times 10^{26}$ | 1995 Dec | 2.6 | 0.9 | 0.5 | -8 | 1.19 | $2.45 \times 10^{10}$ | SS |
| 2113+293 | 0.94 | $3.3 \times 10^{27}$ | 1995 Apr | 2.1 | 0.9 | 0.5 | -1 | 0.87 | $3.66 \times 10^{10}$ | SS |
| 2131-021 | 1.22 | $7.6 \times 10^{26}$ | 1995 Jul | 1.3 | 1.5 | 0.7 | 9 | 0.83 | $9.30 \times 10^{09}$ | SS |
| 2134+004 | 5.51 | $2.9 \times 10^{28}$ | 1995 Jul | 4.3 | 1.3 | 0.6 | 9 | 2.21 | $6.30 \times 10^{10}$ | SS |
| 2136+141 | 1.92 | $1.4 \times 10^{28}$ | 1995 Jul | 2.0 | 1.2 | 0.6 | 14 | 1.45 | $5.21 \times 10^{10}$ | SS |
| 2144+092 | 0.56 | $1.2 \times 10^{27}$ | 1995 Jul | 1.4 | 1.2 | 0.6 | 13 | 0.44 | $9.79 \times 10^{09}$ | SS |
| 2145+067 | 6.52 | $1.1 \times 10^{28}$ | 1997 Mar | 7.9 | 1.1 | 0.6 | 3 | 3.70 | $8.45 \times 10^{10}$ | C |
| 2200+420 | 3.23 | $3.7 \times 10^{25}$ | 1997 Mar | 3.2 | 1.0 | 0.6 | 7 | 1.49 | $2.00 \times 10^{10}$ | SS |
| 2201+315 | 3.10 | $6.1 \times 10^{26}$ | 1997 Mar | 2.4 | 1.0 | 0.5 | 5 | 2.28 | $4.48 \times 10^{10}$ | SS |
| 2209+236 | 0.91 | - | 1995 Apr | 2.3 | 0.9 | 0.5 | 6 | 0.80 | $1.35 \times 10^{10}$ | SS |
| 2223-052 | 3.92 | $1.2 \times 10^{28}$ | 1997 Mar | 2.3 | 1.2 | 0.5 | 5 | 3.28 | $9.93 \times 10^{10}$ | SS |
| 2230+114 | 2.33 | $4.3 \times 10^{27}$ | 1995 Jul | 2.5 | 1.3 | 0.7 | 16 | 0.81 | $1.37 \times 10^{10}$ | SS |
| 2234+282 | 1.21 | $1.4 \times 10^{27}$ | 1995 Apr | 2.1 | 0.9 | 0.5 | 2 | 0.74 | $2.25 \times 10^{10}$ | SS |
| 2243-123 | 2.56 | $2.0 \times 10^{27}$ | 1997 Mar | 1.9 | 1.2 | 0.5 | 0 | 1.85 | $3.80 \times 10^{10}$ | SS |
| 2251+158 | 8.86 | $1.2 \times 10^{28}$ | 1997 Mar | 5.5 | 1.1 | 0.6 | 10 | 4.24 | $9.02 \times 10^{10}$ | SS |
| 2345-167 | 1.60 | $1.1 \times 10^{27}$ | 1997 Mar | 1.8 | 1.2 | 0.5 | 4 | 0.75 | $1.50 \times 10^{10}$ | SS |